**Vacancy-driven non-cubic local structure and magnetic anisotropy tailoring in $Fe_xO$-$Fe_{3-\delta}O_4$ nanocrystals**

Alexandros Lappas,[1,*] George Antonaropoulos,[1,2] Konstantinos Brintakis,[1] Marianna Vasilakaki,[3] Kalliopi N. Trohidou,[3] Vincenzo Iannotti,[4] Giovanni Ausanio,[4] Athanasia Kostopoulou,[1] Milinda Abeykoon,[5] Ian K. Robinson[6,7] and Emil S. Bozin[6]

[1]Institute of Electronic Structure and Laser, Foundation for Research and Technology - Hellas, Vassilika Vouton, 71110 Heraklion, Greece

[2]Department of Chemistry, University of Crete, Voutes, 71003 Heraklion, Greece

[3]Institute of Nanoscience and Nanotechnology, National Center for Scientific Research Demokritos, 15310 Athens, Greece

[4]CNR-SPIN and Department of Physics "E. Pancini", University of Naples Federico II, Piazzale V. Tecchio 80, 80125 Naples, Italy

[5]Photon Sciences Division, National Synchrotron Light Source II, Brookhaven National Laboratory, Upton, 11973 NY, USA

[6]Condensed Matter Physics and Materials Science Department, Brookhaven National Laboratory, Upton, 11973 NY, USA

[7] London Centre for Nanotechnology, University College, London WC1E 6BT, UK

* Correspondence: lappas@iesl.forth.gr; Tel.: +30 2810 391344

**Abstract**

In contrast to bulk materials, nanoscale crystal growth is critically influenced by size- and shape-dependent properties. However, it is challenging to decipher how stoichiometry, in the realm of mixed-valence elements, can act to control physical properties, especially when complex bonding is implicated by short and long-range ordering of structural defects. Here, solution-grown iron-oxide nanocrystals (NCs) of the pilot wüstite system are found to convert into iron-deficient rock-salt and ferro-spinel sub-domains, but attain a surprising tetragonally distorted local structure. Cationic vacancies within chemically uniform NCs are portrayed as the parameter to tweak the underlying properties. These lattice imperfections are shown to produce local exchange-anisotropy fields that reinforce the nanoparticles' magnetization and overcome the influence of finite-size effects. The concept of atomic-scale defect control in subcritical size NCs, aspires to become a pathway to tailor-made properties with improved performance for hyperthermia heating over defect-free NCs.

**Introduction**

Iron oxides are at the research forefront as they encompass important mixed-valent states [1] that impact their physical properties and their technological potential, [2] extending from energy storage devices to catalysts and electrochemical cells. Moreover, since the principal oxidation states (II-IV) of iron carry atomic magnetic moments, spontaneous co-operative magnetic order is stabilized, which offers a highly exploitable modality, extending from spintronics and recording media [3] to the rapidly developing nanobiotechnology. [4] In the latter field of interest, nanoscale magnetic particles for biomedical applications draw significant benefits from the volume-dependence of magnetism, and especially when superparamagnetism (a state not permanently magnetized) is established below a critical particle size. [5] Consequently, there is high demand for controlling the crossover amongst different states of magnetisation in order to improve the particles' magnet-facilitated performance, for making image contrast agents, heat emission "hyperthermia" systems, or even mechanical force nanovectors. [6]

More specifically, radio frequency magnetic heating of single-crystalline nanoparticles (i.e. nanocrystals), is emerging as a novel strategy for activating temperature-sensitive cellular processes, [7] but requires non-toxic biocompatible nanomaterials and understanding how structure-morphology relationships can be used as design parameters. Optimizing hyperthermia efficacy depends on magnetic loss mechanisms attributed to Néel-Brown relaxations, which evolve with the magnetic anisotropy constant ($K$) and saturation magnetization ($M_S$). [8] In practice, relaxation times vary by changing intrinsic nanocrystal factors, including size, [9] shape, [10] and composition [11]. While single magnetic cores of pure iron-metal particles may offer superior heating efficiency, their questionable stability in biological media, [12] has led researchers to develop iron-oxide particles (e.g. ferrites: $Fe_{3-\delta}O_4$, $\gamma$-$Fe_2O_3$ etc) as a versatile and biocompatible class of materials. [13]

The quest for nanocrystals (NCs) that surpass the performance of a single magnetic core is motivated further by the design-concept of controlling the spatial distribution of chemical composition within a single motif, [14] such as core@shell and thin-film heterostructures. Distinct phases grown in a core@shell topology, with contrasting magnetic state constituents (e.g. antiferromagnetic (AFM), ferro- and/or ferri-magnetic (FM/FiM) subdomains), offer a powerful way to tune the nanoscale magnetic properties and boost the particles' hyperthermia response. [10] This rests on the important capability to adjust the particle's anisotropy through the interfacial exchange interactions between the bi-magnetic component phases. [15] The extent to which the emerging exchange-bias ($H_{EB}$) [16] plays a key role is regulated by varying the core-shell volume ratio, surface/interface structure and the composition itself. [17] Thus, optimal design of hyperthermia agents requires the right kind of defect-structure to modulate favorable magnetic relaxations. As we show in this work, these mechanisms can be modeled using Monte Carlo methods.

In this endeavor solution-chemistry methods are widely used to develop size- [18] and shape- [19] controlled ferrite-based nanocrystals, and facilitate their interfacial connectivity on a nanoscale motif, e.g. $Fe@Fe_{3-\delta}O_4$, [20] $CoO@CoFe_2O_4$, [21] $FePt@MFe_2O_4$ (M= Mn, Fe, Co, Zn), [22] etc.. Within these systems, unforeseen magnetic properties are occasionally reported, [23], [24], [25], [26] especially when thermodynamically metastable phases such as wüstite [27] or novel interfaces are introduced during the nanoscale particle nucleation and growth. For example, during the oxidative conversion of AFM wüstite (Fe$_x$O, Fig. 1a), invoking clustering of interstitials and vacancies (Fig. 1b, c), [28] into a FiM spinel magnetite (Fe$_{3-\delta}$O$_4$, Fig. 1d), the internal structure of the as-made core@shell Fe$_x$O@Fe$_3$O$_4$ NCs evolves with the composition gradients influenced by stresses. [29] While a reduction in the AFM@FM interface area and core anisotropy, lowers H$_{EB}$, [30] antiphase boundaries (APBs) in the structure, are found to raise the particle anisotropy. Surprisingly, this paves the way to non-zero H$_{EB}$ even in the fully oxidized, Fe$_{3-\delta}$O$_4$-like derivatives of wüstite. [25] It is interesting to note that APB defects are a favored low-energy growth pathway for Fe$_3$O$_4$-films [31] that modify the exchange interactions and give rise to anomalous magnetic behavior with important applications. [32], [33] These peculiar performances, for NCs in the critical size range of 20-30 nm, were correlated with perturbations of the periodic potential of the iron atomic coordination by lattice defects, which are hard to waive out in post-synthesis treatments, but are preventable by strategic redox tuning of the Fe-valence during the reaction itself. [34]

The preceding discussion demonstrates that the properties of otherwise single-crystalline ferrite particles are influenced by a significant fraction of atoms at the various particle "limits", the surfaces and internal interfaces. [35], [36], [37] This necessitates characterization of the detailed atomic arrangements within the Fe$_x$O-Fe$_3$O$_4$ phase-space available to the chemical synthesis methods. Effectively, research efforts focus on solving a multiple length scale problem, [38] by combining surface (e.g. electron microscopy) and bulk (e.g. powder diffraction) sensitive probes. Total scattering experiments though, coupled to atomic pair distribution function (PDF) analysis, can reach beyond such limitations. [39] Our present work reports on the added value of the PDF method, which, in contrast to near-neighbor probes like EXAFS and NMR, is able to probe a relatively wide field of view (~10 nm) in a single experiment by reporting how local (nanoscale) distortions are correlated throughout the structure.

While chemical phase and stoichiometry ultimately control nanoscale magnetic properties, the rational choice of the critical particle size, with optimal magnetic anisotropy, [40] is ~20 nm and determines the applications. [39], [41] In practice though, thermodynamic and kinetic parameters at nanoscale surfaces and interfaces are expected to trigger nanoscale crystal growth via energetically favorable structural-defect pathways. This serves to render the control of defects as an extra tuning knob. Here, we investigate the nature of structural defects established during the course of the spontaneous, oxidative conversion of wüstite into Fe$_x$O-Fe$_3$O$_4$ NCs. We focus on a series of NCs with increasing particle size in the range 8-18 nm, which differ structurally and morphologically. Our results establish a quantitative relationship between favorable vacancy-induced disorder and tailored magnetic properties, a potentially important tweaking factor at subcritical particle sizes (<20 nm). Monte Carlo simulations demonstrate broader implications for the right kind of defect structure to mediate the magnetic loss mechanisms in favor of efficient energy transfer into heat even for 10 nm NCs.

**Results**

**1   Structural insights**

*1.1   Single-particle local structure*

Four nanoparticle samples were made available for this study, with specimen of spherical shape, entailing diameters of 8.1 ± 0.6 nm and 15.4 ± 1.3 nm and of cubic shape, obtaining edge-lengths of 12.3 ± 0.7 nm and 17.7 ± 1.6 nm (Fig. S1). Henceforth, these are called S8, S15, C12 and C18, where 'S' stands for spherical and 'C' for cubic morphologies. Moreover, high-resolution transmission electron microscopy (HRTEM) (Fig. 2a-d) suggests that the smaller S8 and C12 NPs entail a domain of a single-phase material, but the bimodal contrast in the larger S15 and C18 points that above a particle size of ~12 nm two nanodomains are attained. The coexistence of dark and light contrast features in the S15 and C18 could be justified assuming that two chemical phases, of varying electron diffracting power, share the same nanoparticle volume.   Crystallographic image processing by Fast Fourier transform (FFT) analysis of the relevant HRTEM images results in their corresponding (spot) electron diffraction patterns (Fig. 2e-h). The unequivocal indexing of reflections in the S8 and C12 specimen may suggest that these adopt the magnetite type of phase (Fig. 2e, f). On the other hand, a similar conclusion cannot be made after the evaluation of the FFT patterns of the apparently bimodal contrast S15 and C18 specimen as the cubic spinel and rock salt reflections (Fig. 2g, h) appear to be resolution limited. The behavior appears in line with the tendency of bulk wüstite for oxidative conversion, [28], [42] and infers a process analogous to its elimination from ~23 nm core@shell $Fe_xO@Fe_{3-\delta}O_4$ nanoparticles. [25]

The long and short-range modulations of the selected (hkl) atomic planes could become more easily isolated with direct space images derived from the inverse FFT synthesis of chosen families of reflections. We find that the (400) (or (200) for the core@shell S15 and C18 NPs) family generates perfect atomic planes (Fig. S2.1 to Fig. S2.4) and their spacing could be attributed to the $Fe_{3-\delta}O_4$ spinel (or $Fe_{1-x}O$ rock-salt) type of structure. However, the (220) family (Fig. 2i-l), deviates from being faultless (Fig. S3). Geometric Phase Analysis (GPA) [43] of the (220) reflections depicts single-colored regions (Fig. 2m-p) that are internally homogeneous, showing no obvious inner side distortions. The results illustrate homogeneous structure for the S8 particle (Fig. 2m), but increasing degree of lattice heterogeneity with size (Fig. 2n-p). GPA has pointed before to a 5% crystal lattice deformation from (220) planes of ~23 nm $Fe_xO@Fe_{3-\delta}O_4$ nanocubes, while that for the (400) (or (200) for a rock salt) family amounted to only ~1%. [25]

In summary, the evaluation of the number (Nr) of defects involving the (220) planes, implies that NPs of spherical morphology (S8, S15) carry a larger number of defects in their volume than the cubic ones, and those of smaller size (S8) are somewhat more susceptible to lattice faults than those of larger size (Fig. S4; §S3 and Table S1). Effectively, such distorted (220) atomic planes (Fig. 1d) impose tensile lattice strain as the overall prevailing effect that is pronounced for the spherical morphology (cf. from 4-5% in the latter, drops down to 1-2% in cubic shape NPs; §S4 and Table S2).

*1.2   Ensemble-average local structure*

With the aim to go beyond the HRTEM findings and acquire quantitative phase-specific structural information from a large ensemble, we measured the synchrotron X-ray PDF of selected nanoparticle specimens and compared them to the bulk magnetite (Fig. 3). As moderate Q-space resolution of the experimental setup used limits the PDF field of view in the r-space (r> 5-10 nm length-scales met in bulk), our analysis focuses primarily on the low-r region in the atomic PDF (r= 1- 10 Å). This practically allows describing the local structure through insights on bonding and lattice distortions within the $Fe_{3-\delta}O_4$ and $Fe_xO$ unit cells.

### 1.2.1 Local structure distortions

Bulk magnetite was taken as a reference system against which subtle distortions of the atomic lattice planes in the nanoparticles could be identified. Initial refinements were performed with the simplified, normal spinel, cubic configuration of magnetite, $(Fe^{3+})_8[Fe^{3+},Fe^{2+}]_{16}O_{32}$, [44] where the round brackets represent tetrahedral (Td) and the square brackets octahedral (Oh) coordination by oxygen crystallographic sites (i.e. model #1; this assumes no $Fe^{2+}/Fe^{3+}$ inversion between Td/Oh sites - Table S3, §S5). The atomic PDF in the low-r region for the bulk sample at 300 K is described well by this textbook model ($R_w$= 8.2%). A somewhat lower quality of the fit at 80 K (Fig. 4a; $R_w$= 8.9%) may infer some sensitivity to acentric distortions, beyond the r-space resolution of our measurements, caused by the Verwey transition of bulk magnetite. [45]

The same textbook model (#1) was then utilized to model the xPDF data for nanoparticles of variable size and morphology. What is particularly striking is that between 300 K and down to 80 K, this cubic model, systematically fails to fit the peak at r~ 3 Å (sample S8: Fig. 4b, $R_w$= 18.1%; sample C12: Fig. S6a, $R_w$= 11.9%). Importantly, this corresponds to the closest distance between the iron atoms that are octahedrally coordinated by oxygen in the structure of magnetite (forming a pyrochlore-type sublattice, Fig. 1d). Moreover, the nearest distance of a pair of Fe-Fe in wüstite is also just above 3 Å. It may be expected then that if a detectably large volume fraction of wüstite phase is present in the NPs (cf. HRTEM for S15) then the intensity of the peak at r~ 3 Å should be increased. In an effort to explore this, a two-phase cubic rock-salt and spinel ($FeO - Fe_3O_4$) model was employed. This analysis showed that either in smaller nanocrystals (S8 and C12) or larger NPs (S15), the cubic symmetry model (Fig. 4c; $R_w$= 13.6%), is somewhat misplaced with respect to the 3 Å observed radial distance distribution. Per present analysis and assessments of fits over broader r-ranges it appears that these nanoscale samples are highly non-uniform in terms of defects (*vide-infra*), and their structure is not quite cubic at high-r, and is neither so locally too.

For these reasons, the possibility of the nanoparticle local structure deviating from the ideal cubic lattice configuration has been factored-in. However, the modest Q-space resolution for resolving symmetry-lowering configurations beyond the local scale, and in turn limited PDF field of view, led us to utilize approximations implemented by space groups of higher symmetry. The exemplary case of maghemite ($\gamma$-$Fe_2O_3$) drove this effort as it can be considered as $Fe^{2+}$-deficient magnetite, $(Fe)_8[Fe_{1\frac{1}{3}}\langle\ \rangle_{2\frac{2}{3}},Fe_{12}]O_{32}$ ($Fe_{Oh}$-vacancies represented by the angular $\langle\ \rangle$ brackets) [46], [47]. Under this defect-based scheme randomly distributed vacancies result in an fcc lattice (model #1), but when their ordering is favored, either a primitive cubic ($P4_332$ symmetry, model #2) [48], [49] structural variant could be stabilized, or a symmetry-lowering lattice distortion is triggered ($P4_32_12$ tetragonal symmetry, model #3) [50], [51].

Amongst these, only model #3 (Table S3, §S5) made a marked improvement in the description of the PDF peak positions from room temperature down to 80 K (Fig. 4d, Fig. S6b, Fig. S7c). The smaller NPs (≤12 nm) were assessed by PDF (Figure S5, §S5) to be 100% tetragonal at the local level, while the larger one (S15) entailed a 22.5%:77.5% [rock salt]:[tetragonal] share of volume fractions. The PDF intensities were accurately described when the Fe-site occupancies ($\eta$) were refined. The outcome indicates that nanoparticle samples of spherical morphology display the least occupied Fe-sites, namely: $\eta$-S8 < $\eta$-S15< $\eta$-C12 [i.e. 16.4(1)< 18.2(1)< 20.0(1), out of 24 Fe-atoms / unit cell; Fig. S8, §S8]. Along this trend comes the expansion of the in-plane (a-b) lattice dimensions and the contraction of the c-axis, which suggest a more pronounced tetragonal compression for the spherical (S8, S15 c/a~ 0.975) rather than the cubic (C12, c/a~ 0.986) morphology NPs. Details on the atomic PDF analysis, assessment of the outcomes and a summary of the derived parameters (Table S4) are presented in the supplemental material (§S5).

Though the local symmetry change may be seen as a manifestation of the coupling between elastic and exchange energy terms, [52] which are likely optimized by the apparently larger strain in the nanospheres, subtle crystalline electric field effects due to a local tetragonal Jahn-Teller distortion, lifting the orbital degeneracy in $Fe^{2+}$ ($3d^6$), [53] cannot be either resolved or ruled-out by the structural xPDF probe on this occasion.   Moreover, while we cannot completely discard the contribution of valence-swap induced distortion [with a fraction of $Fe^{2+}$ atoms occupying $Fe^{3+}$ sites and vice versa {cf. $(Fe^{3+})_8[Fe^{3+},Fe^{2+}]_{16}O_{32}$}], [54] we emphasize that vacancy-driven effects are a plausibility as the relative G(r) peak intensities of $Fe_{Oh}$-$Fe_{Oh}$ ($r$~ 3.0 Å) and $Fe_{Oh}$-$Fe_{Td}$ ($r$~ 3.5 Å) change appreciably (*vide-infra*), whereas the one-electron difference between the $Fe^{2+}$ and $Fe^{3+}$ electronic configurations would effectively be unobservable in the xPDF peak heights. It is therefore inconceivable that observed dramatic changes in relative PDF peak intensities originate from inverted-spinel-like electronic configurations.

Recapitulating the above PDF analysis, it is important to recognize the sensitive nature of the NPs, and especially those with spherical shape, in the local structure non-stoichiometry. Fe-vacancies stabilize a tetragonally distorted local structure, inferring a relation to the large number of defects in the crystal volume, as implicated by the HRTEM based analysis as well. Our results, albeit based on just two single-phase specimens, suggest enhanced tetragonality for spherical nanoparticle morphology.

### 1.2.2   Where the defects are located

The question as of how structural vacancies relate to different cation lattice sites is now tackled by comparing the observed, normalized G(r) patterns of the NPs against the bulk magnetite (Fig. 5a). Two types of radial distance populations, represented by the G(r) peak-intensity, are considered. (a) $Fe_{Oh}$-$Fe_{Oh}$ separations ($r$~ 3.0 Å): When the single-phase NPs peak-intensity maximum is compared to that in the bulk stoichiometric $Fe_3O_4$ a measure of the presence of Oh vacancies is witnessed. With the ratio being ~0.8 and ~0.9, for the S8 and C12 NPs, respectively, more Oh-Fe vacancies are supported in the nanospheres. For the larger NPs (S15) the enhanced G(r) suggests an increased rock-salt type of phase in their volume (§1.2.1). (b) $Fe_{Oh}$-$Fe_{Td}$ separations ($r$~ 3.5 Å): The progressive diminution of the G(r) peak-intensity advocates also that the spherical nanoscale morphology (S8, S15) adopts noticeably more empty lattice sites than the cubic one (C12), conferring that their abundance is a shape-dependent phenomenon. [10]

With the purpose of evaluating further if the vacancies have a site-specific preference, xPDF patterns were simulated, while the ratio of Fe-vacancy population at the Oh and Td sites was varied. The trend is similar assuming either the symmetry-lowering model #3 (Fig. 5b) or the cubic spinel model #1 (Fig. S9b, c; §S9). The progressive intensity diminution at $r \sim 3.5$ Å, when raw and simulated patterns are compared, corroborates to a significant volume of vacancies also at the Td Fe-sites (~10-20%; Fig. S8), inferring a limited length of structural coherence. [55] This is in contrast to bulk spinel samples where intrinsic Oh Fe-vacancies mediate the structure and properties. [56] The significant content of vacancy distribution resolved by xPDF may infer emerging strains/stresses (Fig. S10 and §S10), reminding those in $Fe_xO@Fe_3O_4$ nanocubes probed by single-particle local structure techniques. [29], [25]

Overall, PDF indicates that during the self-passivation of wüstite, smaller NPs are single-phase, while larger ones attain a two-phase character. This size-mediated phase evolution is in agreement with the HRTEM findings, which also indicate that particles of spherical shape accommodate larger number of defects in their volume [cf. xPDF refined Fe-site occupancy $\eta$-S8 < $\eta$-S15< $\eta$-C12 [i.e. 16.4(1)< 18.2(1)< 20.0(1), where $\eta$= 24 Fe-atoms / unit cell of the bulk cubic spinel; Fig. S8]. Besides, PDF uncovers that in addition to vacancies commonly found at the Oh sites, Td-Fe is largely absent, fostering local tetragonal lattice distortions. These observations trigger questions about the impact of vacancies on the observed properties.

## 2 Nanometer scale effects on properties

### 2.1 Magnetic behavior

In view of the nanoparticles' deviation from perfect structural ordering their magnetic behavior is evaluated here because of its direct relevance to hyperthermia applications. A broad maximum in the *dc* magnetic susceptibility, $\chi(T)$, with an irreversibility between ZFC/FC curves (Fig. 6a-c; Fig. S11a), marks a characteristic temperature, $T_B$, that separates the superparamagnetic state from the blocked state. [5] In addition, the $\chi(T)$ for the core@shell NPs (S15; Fig. 6c) points to a sudden drop due to the paramagnetic to AFM transition ($T_N$) in wüstite and a subtle anomaly resembling the Verwey transition ($T_V$) of bulk magnetite. [30], [35], [36] Furthermore, the evolutions of the hysteresis loop characteristics (§S11, Table S5; Fig. S11b-c) suggest that a processes beyond the coherent reversal of *M* is involved. In this, $M(H)$ experiments under field-cooling ($H_{cool}$= 50 kOe; Fig. 6d-f, Fig. S12, S13), support the development of a macroscopic, exchange-bias field ($H_{EB}$) resulting from interfacial interactions [16]. The quick rise of $H_{EB}$ for the S15 NPs, against the single-phase S8 and C12 (Fig. 7, Fig. S14, §S14), implies competing exchange interactions due to different kinds of interfaces.

In addition, all the NPs present a discontinuous step-like variation of the magnetization near zero field (Fig. 6d-f, Fig. S13). The two-switching field distributions, marked by maxima in dM/dH, resemble the inhomogeneous magnetic behavior arising from coexisting magnetic components of contrasting $H_{CS}$ (e.g. of mixture of particle sizes [57] or compositions [58]). Here though, the HRTEM study of the $Fe_xO$-$Fe_3O_4$ NPs confers their single crystal character and narrow particle size distribution (Fig. S1). However, xPDF resolves Fe-site vacancies [$V$= ($\eta$-S8 or $\eta$-S15 or $\eta$-C12): $\eta$ is the modeling-derived content per unit cell volume, $\eta$= 24 Fe-atoms / cell of bulk cubic spinel], with $V$-S8 (33%) > $V$-S15 (25%) > $V$-C12 (17%). The existence of these defects seems to establish a spatial variation of the composition at the local level that 'turns on' the observed inhomogeneous magnetism.

### 2.2   Coupling of structural defects to magnetism

To shed light on how atomic-scale defects (e.g. Td Fe-lattice sites vacancies and local tetragonal distortions) couple to magnetism we utilized Monte Carlo (MC) simulations. For this purpose, our systems were approximated by a microscopic "core-surface" model, [59], [60] however, with random defects introduced in the nanoparticle structure. These defects were described as weakly coupled FM pairs of spins with strong anisotropy, inferred from the symmetry-lowering local structure (§1.2.1) of the NPs. Their soft FiM character [61] was chosen to resemble that of $V_4$-Td clusters of defect-units [due to coalescence of 4 $Fe_{Oh}$-vacancies ($V$) around an $Fe^{3+}_{Td}$-interstitial] (Fig. 1b-c) [28] out of which spinel magnetite has been claimed to nucleate during the oxidative conversion of $Fe_xO$. [62] Assuming that the fraction of the FM pairs of spins (pinning bonds) is a tunable particle parameter, associated with the xPDF-derived Fe-vacancies, the evolution of (a) the low-field jump ($\Delta M/M_S$, estimates how many magnetic moments are switched) and (b) the exchange-bias ($H_{EB}$) have been quantified (Fig. 8).

With the purpose to assess the former, $\Delta M/M_S$, we note that the surface is a large part for the smaller nanoparticles, i.e. ~50% of the S8 and ~35% of the C12 specimens, and both may assume a two-phase magnetic nature. The $\Delta M/M_S$ changes may then be a manifestation of the character stemming from the exchange-coupled hard and soft ferro/ferri- magnetic phases in the two nanodomains. [63] As the number of pinning bonds (vacancy-driven) increases in the main body of these NPs, MC calculations point that the defects' strong random anisotropy leads to exchange randomness rendering their two-phase magnetic nature more disordered. This leads to less prominent, $\Delta M/M_S$ changes in the hysteresis loops (Fig. 8). Considering the evaluation of the latter, we note that the defect-induced spin disorder within the core-like sub-domains reduces the overall magnetization (Fig. S12), but supports extra pinning centers that could foster a density of uncompensated interfacial spins, impeding easy coherent reversal [64] of the surrounding ferrimagnetic moments (Fig. S13). In line with this are the MC simulations which indicate that in small S8 and C12 NPs without defects, exchange-bias is absent, but when the perturbation of the periodic potential of the iron coordination by lattice defects is enabled, the increasing number pinning bonds in the core result in net exchange bias (Fig. 8).

The behavior is depicted in the magnetic moment snap-shots of defected (Fig. 9a-c) versus defect-free (Fig. 9d-f) NPs. While for non-defected NPs the soft anisotropic core follows the applied field reversal, the existence of defects generates localized antiparallel spin components, which couple to neighboring ferrimagnetic spins at the atomic scale interface, and promote the canting of the core towards the xy-plane (snap-shots of the spin ensemble under a full M-H loop, are compiled in Fig. S15, while the mean moment orientation is shown in Fig. S16). In this way, the nanospheres' more defected internal structure generates adequate conditions that endorse exchange-coupling so that $H_{EB}$-S8> $H_{EB}$-C12 (while $H_C$-S8< $H_C$-C12, due to differences in the NPs' magnetic volumes). As magnetization states at the different kinds of interfaces are adjusted by H$_{cool}$, for a matter of consistency, it is worth pointing out that MC simulations (Fig. S17) reproduce fairly well the experimental (Fig. 7) evolution of the hysteresis loop parameters (H$_{EB}$, H$_C$, $\Delta M/M_S$) even for the larger S15 NPs (§S17).

The defect-rich NPs discussed here broaden the picture that growth-approach mediated subtle, structural microscopic factors, [65], [24], [33] foster local-scale anisotropy that facilitates exchange bias in otherwise phase-pure, monocrystalline NCs.

### 2.3 Defect-driven magnetic heating

The observed evolutions may be related to structural and morphological variations between NPs exhibiting differences in size, surface anisotropy and exchange anisotropy. Ferrite nanocrystals, in magnetically-mediated biomedical applications, draw their versatility from the critical particle size (~20 nm), [39] a factor which varies with the magnetic anisotropy. [41] In the present work we have demonstrated that surface atoms can respond differently than the core ones, a prominent effect for the smaller fully oxidized derivatives (≤12 nm) of the Fe$_x$O-Fe$_{3-\delta}$O$_4$ NCs. Although defect-elimination during synthesis can yield nanomagnetic agents (≥20 nm) with enhanced, concurrent diagnostic imaging and thermoresponsive performances, [34] structural defects at sub-critical particle sizes appear to offer a different exploitable pathway, compatible though with the biological limits (e.g. set by toxicity and patient discomfort). [66] Here, vacancies in self-passivated iron oxides of subcritical size (≤12 nm), act as pinning centers that favour the competition of exchange interactions, thus fostering local anisotropy enhancement. Benefits from the NPs' extended anisotropic properties may raise their application potential, as for example to afford heat generation beyond the bare susceptibility losses (Néel-Brown relaxation) mediated by finite-size effects alone (§S18). [40]

The MC simulations support that for small defected NPs (S8), where the effective anisotropy increases 5 times compared to the defect-free ones, the heat dissipation (specific absorption rate = SAR) is raised almost ten-fold, i.e. ~450 W/g *vs.* ~50 W/g, at 500 kHz under 37.3 kA/m (Fig. 9g). For comparison, it is worth noting that experimentally, defect-free 9 nm Fe$_3$O$_4$ NPs, exhibit a SAR of 152 W/g. [14] It can be envisaged then that nanoparticle self-passivation leads to adequate perturbation of the periodic potential of the iron coordination by vacant lattice sites that in turn tweak the core to surface magnetic anisotropy ratio. Whether this might be an avenue to boost the thermal energy transfer at subcritical particle sizes (≤10 nm) by synergistic relaxation processes, warrants further exploration. Even in the absence of extended buried AFM/FM interfaces, [14] where beneficial exchange parameters attain efficient heat sources, smaller, heterogeneous nanocrystals (S8, C12) may be uncovered as useful heat-therapeutic agents.

## 3 Conclusions

The self-passivation of nanoscale wüstite (Fe$_x$O) has been investigated as a testing ground to explore the consequences of thermodynamically unstable interfaces formed during the nucleation and growth of nanostructured iron-oxides. As conventional oxidation evolves, grown component phases give rise to single-crystal nanoscale entities with sub-domain Fe$_x$O-Fe$_{3-\delta}$O$_4$ interfacial connectivity. However, when the synthesis parameters are varied to attain smaller-size nanocrystals (≤12 nm) a spinel-like phase is nucleated alone. The compositional and structural complexity of these iron-oxide nanostructures is witnessed by single-particle transmission electron microscopy and complementary, phase-specific structural information, attained from a large ensemble by high-energy synchrotron X-ray total scattering experiments. These uncover that defects (with a significant volume residing at tetrahedral Fe-sites) alleviate a surprising tetragonal lattice compression in the spinel-like phases. The defects entail structural vacancies with an increased number density when the nanoscale morphology changes from cubic to spherical and the particle size shrinks. Moreover, magnetometry and Monte Carlo calculations show that the nanostructures' heterogeneous character reflects in a core-surface type of spin-configuration that favors two switching field distributions. Larger core@shell $Fe_xO-Fe_{3-\delta}O_4$ nanocrystals support exchange-bias due to the coupling across a common interface of spatially extended sub-domains of AFM and FM nature. Exchange-bias is unexpectedly evident, though much reduced, in smaller-size (≤12 nm), fully oxidized particles, due to the existence of the defected internal structure which generates localized antiparallel spin components, and uncompensated spin density at atomic-scale interfaces. The latter, in line with the non-cubic local symmetry of the nucleated spinel phases, expresses the influence of local anisotropy fields, which apparently deviate from the easy-axis symmetry met in common iron-oxides, favoring canting into the xy-plane. The results corroborate that size-dependent evolution of the metal-cation valence state, produces pinning defects which promote the competition of the exchange interactions at subcritical sizes (< 20 nm). The concept raised here points that atomic-scale defect control in small particles (~10 nm), typically hampered by the superparamagnetic limit, may act in favor of anisotropic properties for improved magnetism-engineered functionalities (cf. heating agents and thermo-responsive cellular processes).

## Methods

### 4.1 Materials

All reagents were used as received without further purification. Oleic acid (technical grade, 90%), absolute ethanol (≥ 98%), octadecene (technical grade, 90%), hexane (ACS reagent, ≥99%) and sodium oleate powder (82%) were purchased from Sigma Aldrich. Iron (III) chloride ($FeCl_3 \cdot 6H_2O$, ACS Reagent) was purchased from Merck. Iron (III) acetylacetonate was purchased from Alfa Aesar and oleylamine 80-90% was purchased from Acros Organics.

### 4.2 Syntheses protocols

Colloidal syntheses were carried out in 100 mL round-bottom three-neck flasks connected via a reflux condenser to a standard Schlenk line setup, equipped with immersion temperature probes and digitally controlled heating mantles. The reactants were stored under anaerobic conditions in an Ar-filled glovebox (MBRAUN, UNILab), containing <1ppm $O_2$ and $H_2O$.

 A gas mixture of 5% $H_2$/Ar has been used as a protective/reductive atmosphere. The reductive atmosphere can help to maintain the $Fe_xO$ (wüstite) phase instead of the oxidized $Fe_3O_4$ (magnetite) and/or $\gamma$-$Fe_2O_3$ (maghemite) forms. Previous studies have shown that $Fe_xO$ crystallites formed under these synthetic conditions become rapidly oxidized after removing the reducing agent and exposing them to ambient air. [67] Very small particles tend to be fully oxidized to the spinel structure. A minimum diameter over ~13 nm is needed for the nanoparticle to maintain its core@shell structure.

### 4.2.1 Preparation of iron oleate precursor

Iron (III) - oleate was prepared before each nanoparticle synthesis and used as an iron precursor subsequently. Special care was taken to protect it from the light. The metal oleate precursor is formed by the decomposition of $FeCl_3 \cdot 6H_2O$ in the presence of sodium oleate at 60 °C, based on a slightly modified literature protocol. [68] 16 mmol of $FeCl_3 \cdot 6H_2O$ salt and 48 mmol of sodium oleate were dissolved in a mixture of solvents in a round bottom flask. 56 mL hexane, 32 mL ethanol and 24 mL distilled water were used as solvents. The mixture was heated to 60-65 °C under Ar atmosphere for 4 hours and then let to cool down to room temperature. The organic phase containing the metal oleate complex was separated from the aqueous phase using a separatory funnel, then washed with ~30 mL distilled water and separated again. This process was repeated 4 times and at the end the metal organic complex was dried under stirring and mild heating for several hours, resulting in a viscous dark red oleate. The final product was stored in a dark place to protect it from light. Some mild heating to ensure its fluidity may be needed just before its use for each nanoparticle synthesis. The successful fabrication of the ferric oleate complex has been identified by the FTIR data (Fig. S19). [69]

### 4.2.2 Synthesis of iron-oxide nanoparticles

The nanoparticles (NPs) were synthesized by employing modified literature protocols [69], [35], [70], [18] aiming to produce the wüstite type of oxide ($Fe_{1-x}O$). In a typical synthesis 2-7 mmol of iron oleate were dissolved in octadecene in a flask under a reductive atmosphere. Oleic acid was used as surfactant and protective ligand, in a proportion 1:2 with respect to the iron precursor. The amounts of reactants were tuned so that a final Fe-oleate molar ratio of 0.2 mol/kg solution was achieved. The synthesis protocol includes three major steps. First, a degassing step at 100 °C for 60 min under vacuum is required for the complete removal of any water and oxygen residues. Then the mixture is heated to 220 °C with a heating rate of ~10 °C·min$^{-1}$. At this temperature that lasts for 60 min the so-called nucleation step allows for the crystal seeds [92] to be generated for the successive formation of the nanocrystals (NCs). At the final stage, the mixture is heated to 320 °C where the nanocrystals' growth takes place. At the end of the synthesis the colloidal mixture was left to cool down at room temperature and the NCs were precipitated upon ethanol addition. They were separated by centrifugation at 6000 rpm for 5 min, re-dispersed in hexane and then centrifuged once more after adding ethanol in a 1:1 ratio with respect to the hexane. The process was repeated two more times at a centrifugation speed of 1000 rpm. Addition of sodium oleate was proved to promote the formation of cubic NPs, as proposed in earlier studies. [70] We found that a metal precursor to sodium oleate ratio of 8:1 to 5:1 is adequate enough to realize such a shape transformation. Minor variations in this two-step heating protocol allows the tuning of the particle's size and the control of their size distribution. [18], [71] The protocol gives rise to NPs with diameters up to 20 nm, with size-control attained by modifying the growth time (40< t< 90 min) at the final stage. An extended stay here produces even larger NPs, but beyond 60 min smaller-size particles are afforded, likely due to a ripening mechanism. Four iron-oxide colloidal nanostructures, with varying size and morphological features (*vide-infra*), S8, C12, S15, C18 (S: spherical, C: cubic) were finally grown and stored as colloidal dispersions in ~4 mL hexane in septa-sealed vials.

### *4.3 Characterization techniques*

#### 4.3.1 High Resolution Transmission Electron Microscopy (HRTEM)

Low magnification and high resolution Transmission Electron Microscopy (TEM) images were recorded, using a $LaB_6$ JEOL 2100 electron microscope operating at an accelerating voltage of 200 kV. All the images were recorded by the Gatan ORIUS$^{TM}$ SC 1000 CCD camera. For the purposes of the TEM analysis, a drop of a diluted colloidal nanoparticle solution was deposited onto a carbon-coated copper TEM grid and then the hexane was allowed to evaporate. In order to estimate the average size, statistical analysis was carried out on several low magnification TEM images, with the help of the dedicated software ImageJ. [72] The structural features of the nanoparticles were studied by two-dimensional (2D) fast Fourier transform (FFT) images acquired and analyzed by ImageJ.

With the purpose to highlight the defect structure of the nanoparticles we employed the Geometric Phase Analysis (GPA) method of Hÿtch et al.. [43] The HRTEM images were Fourier transformed and a region around one of the {220} peaks was selected with a circular window. This diffraction pattern was re-centered and inverse Fourier transformed to provide a real-space image whose phase (Fig. 2m-p), represents the projection (onto the {220} reflection chosen) of the lattice shift of that particular region of the crystal relative to the average lattice. The single-colored regions are internally homogeneous, showing no obvious internal distortions, but have different phase shifts from their neighbors.

#### 4.3.2 X-ray Pair Distribution Function (xPDF)

X-ray synchrotron-based PDF data were acquired at the 28-ID-1 beamline of the National Synchrotron Light Source II, Brookhaven National Laboratory. Each nanoparticle powder was encapsulated in a $\varnothing$1.0 mm kapton capillary, sealed at both ends with epoxy glue. The 28-ID-2 in PDF mode uses a *Perkin-Elmer* 2D image plate detector for fast data acquisition, but of relatively modest Q-space resolution which, in turn, limits the PDF field of view in r-space. Data were collected between 80< T< 400 K, making use of the beamline's liquid nitrogen cryostream (Oxford Cryosystems 700), with incident X-ray energy of 68 keV. Bulk magnetite powder ($Fe_3O_4$) was utilized as a reference.

The atomic PDF [39] gives information about the number of atoms in a spherical shell of unit thickness at a distance *r* from a reference atom, and is defined as:

$$G(r) = 4\pi r[\rho(r) - \rho_0] \tag{1}$$

where $\rho_o$ is the average number density, $\rho(r)$ is the atomic pair-density, and *r* represents radial distance. The raw 2D experimental data are then converted to 1D patterns of intensity versus momentum transfer, Q (= $4\pi \sin\theta/\lambda$), which are further reduced and corrected using standard protocols, and then finally Fourier transformed to obtain G(r):

$$G(r) = (2/\pi) \int_{Q_{min}}^{Q_{max}} Q[S(Q) - 1] \sin(Q r) dQ \tag{2}$$

S(Q) is the properly corrected and normalized powder diffraction intensity measured from $Q_{min}$ to $Q_{max}$. The experimental PDF, G(r), can subsequently be modeled by calculating the following quantity directly from a presumed structural model:

$$G(r) = [\frac{1}{r}\sum_{ij}\frac{f_i f_j}{<f>^2}\delta(r - r_{ij})] - 4\pi r \rho_0 \qquad (3)$$

here, f stands for the X-ray atomic form factors evaluated at Q= 0, $r_{ij}$ is the distance separating the i-th and j-th atoms, and the sums are over all the atoms in the sample. In the 28-ID-2 experiments, elemental Ni powder was measured as the standard to determine parameters, such as $Q_{damp}$ and $Q_{broad}$, to account for the instrument resolution effects. Experimental PDFs, based on modest Q-space resolution and a $Q_{max}$= 25 Å$^{-1}$ raw powder diffraction data, were fitted with structural models using the PDFgui [73] software suite.

### 4.3.3   Magnetic measurements

The magnetic characterization was conducted using a vibrating sample magnetometer (VSM, Oxford Instruments, Maglab 9T), operating at a vibration frequency of 55 Hz. The measurements of the temperature dependent magnetization, M(T), were carried out at 50 Oe at a fixed temperature rate of 1 K min$^{-1}$ after either zero-field cooling (ZFC) or field cooling (FC) in 50 Oe from 300 K to 5 K. Selected M(T) measurements were also carried out at a different applied field (100 Oe). Hysteresis loops, M(H), were obtained at room temperature and at 5 K by sweeping the applied field from +50 kOe to -50 kOe and back to +50 kOe after cooling the sample from 300 K to 5 K under ZFC or an applied field $0 < H_{cool} \leq 50$ kOe (FC). In the FC procedure, once the measuring temperature was reached, the field was increased from $H_{cool}$ to H = 50 kOe and the measurement of the loop was pursued. Due to a small remnant field, common in superconducting magnets, the values of the coercive field (Table S5) have been corrected, as [magnetometer reported field] + [field error] = [real magnetic field at the sample]. The offset error in the magnetic field (max 60 Oe) was estimated through the "Field Error vs. Charge Field" calibration chart of the magnetometer. Although such an amendment does not propagate to the extracted exchange-bias values, it has been taken into account to the low-field demagnetization ($\Delta M/M_S$), and the ratio of remanence against saturation ($M_r/M_S$) (Figure 7). In addition, when data are recorded under a magnetic field sweep, a synchronization error of the measuring electronics can be observed if high sweeping rates, larger than 200 Oe/s, are chosen. To avoid this artifact, we evaluated M(H) loops collected at step mode versus sweeping, at various rates and found that a sweeping rate of 30 Oe/s provides adequate conditions to minimize the error propagation in the measurements.

While precautions were taken to maintain the integrity of the samples, their spontaneous chemical evolution led us to exclude one specimen (C18) from the in-depth discussion where magnetism and structural properties are correlated (§2). This is because while TEM structural investigations were pursued soon after the sample growth, neither xPDF nor magnetic characterizations were readily available (due to remote facility timeline access restrictions) near the initial life-time of the samples. So, while four iron-oxide colloidal nanostructures were grown initially, their self-passivation into $Fe_xO$-$Fe_3O_4$ nanocrystals, led us to discuss magnetic data (Fig. S11) only for three samples (i.e. S8, C12, S15) that allowed a coherent picture of their structure – property relations to be attained.

### 4.4 Monte Carlo simulations

The simulations approximate the nanoparticles (NPs) by a microscopic "core-surface" model. [59], [60] Three nanoparticle model systems were considered, with somewhat different morphological features, as these were probed experimentally in the S8, C12, and S15 specimens. The spins in the NPs were assumed to interact with nearest-neighbor Heisenberg exchange interaction, and at each crystal site they experience a uniaxial anisotropy. [35], [74], [75]  Under an external magnetic field, the energy of the system is calculated as:

$$E = -J_{core} \sum_{i,j \in core} \vec{S}_i \cdot \vec{S}_j - J_{shell} \sum_{i,j \in shell} \vec{S}_i \cdot \vec{S}_j - J_{IF} \sum_{i \in core, j \in shell} \vec{S}_i \cdot \vec{S}_j$$

$$-K_{i \in core} \sum_{i \in core} \left( \vec{S}_i \cdot \hat{e}_i \right)^2 - K_{i \in shell} \sum_{i \in shell} \left( \vec{S}_i \cdot \hat{e}_i \right)^2 - \vec{H} \sum_i \vec{S}_i \qquad (4)$$

Here $\vec{S}_i$ is the atomic spin at site $i$ and $\hat{e}_i$ is the unit vector in the direction of the easy-axis at site $i$. The first, second and third terms give the exchange interaction between spins in the AFM core, in the FiM shell and at the interface between the core and the shell, respectively. The interface includes the last layer of the AFM core and the first layer of the FiM shell. The fourth and fifth terms give the anisotropy energy of the AFM core, $K_C$, and that of the FiM shell, $K_{shell}$, correspondingly; the last term is the Zeeman energy.

Parameters were chosen (§S20) after careful analysis of the experimental magnetic behavior and Monte Carlo (MC) simulations were implemented with the Metropolis algorithm. [76] The hysteresis loops, $M(H)$, were calculated upon a field cooling procedure, starting at a temperature $T= 3.0$ $J_{FM}/k_B$ and down to $T_f = 0.01$ $J_{FM}/k_B$, at a constant rate under a static magnetic field $H_{cool}$, directed along the z-axis. The hysteresis loop shift on the field-axis gave us an estimate of the exchange-field, $H_{EB} = -(H_{right} + H_{left})/2$. The coercive field was defined as $H_C = (H_{right} - H_{left})/2$. $H_{right}$ and $H_{left}$ are the points where the loop intersects the field-axis. The fields H, $H_C$ and $H_{EB}$ are given in dimensionless units of $J_{FM}/g\mu_B$, the temperature $T$ in units $J_{FM}/k_B$ and the anisotropy coupling constants $K$ in units of $J_{FM}$. In this work $10^4$ MC steps per spin (MCSS) were used at each field step for the hysteresis loops and the results were averaged over 60 different samples (namely random numbers).

### Acknowledgements

This research used the beamline 28-ID-1 of the National Synchrotron Light Source II, a U.S. Department of Energy (DOE) User Facility operated by Brookhaven National Laboratory (BNL). Work in the Condensed Matter Physics and Materials Science Department at BNL was supported by the DOE Office of Basic Energy Sciences.  Both activities were supported by the DOE Office of Science under Contract No. DE-SC0012704. We acknowledge partial support of this work by the project "National Research Infrastructure on nanotechnology, advanced materials and micro/nanoelectronics" (MIS 5002772), which is implemented under the "Action for the Strategic Development on the Research and Technological Sector", funded by the Operational Programme "Competitiveness, Entrepreneurship and Innovation" (NSRF 2014-2020) and co-financed by Greece and the European Union (European Regional

Development Fund). AL thanks the Fulbright Foundation - Greece for a Visiting Scholar award (academic year 2016-2017) to conduct research at the Brookhaven National Laboratory, USA.

**Supplemental Material**

Supplementary Information accompanying this paper includes: TEM size-distributions, HRTEM inverse FFT synthesis, Calculated parameters for defects, Size and shape dependent strain, Details of the PDF analysis, xPDF fits for the C12 and S15 nanoparticles, Simulations of the defect-dependent changes in G(r), Size and shape dependent vacancy content, Cell-size T-evolution for S8 and S15, Variable temperature susceptibility data, ZFC and FC hysteresis loops, Magnetic anisotropy, ZFC and FC hysteresis loops characteristics and their first derivative, Cooling-field dependence of *M(H)* for S15, FTIR spectra, Monte Carlo calculations of the *M(H)* parameters, Spin-configuration snap-shots and SAR for defected S8 and defect-free nanoparticles.

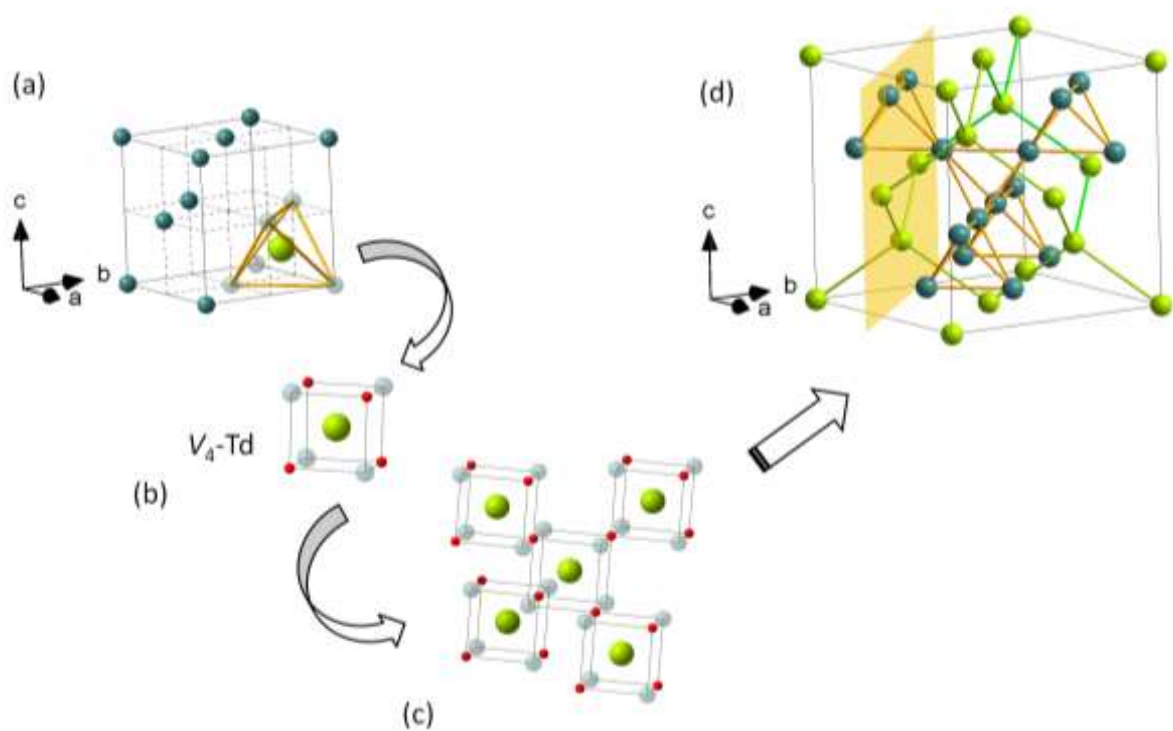

**Figure 1.** Illustrations of the Fe-site arrangement in archetype face-centred cubic crystal types of bulk iron oxides, namely: wüstite ($Fe_xO$) (a), a defect-mediated rock-salt structure, with octahedrally coordinated (by oxygen - not shown for clarity) ferrous ($Fe^{2+}$) sites (Oh; represented by dark green spheres), and interstitial ferric ($Fe^{3+}$) tetrahedral sites (Td; depicted by large light green spheres) in the presence of four vacant ($V$) iron positions (light green). Such units compose $V_4$-Td defect clusters (b) (oxygen shown by red spheres), whose coalescence (c) [28] offers a likely pathway towards the nucleation of $Fe_{3-\delta}O_4$ (magnetite) (d). The latter represents the ferro-spinel family, with Fe-atoms in both types of coordination environment, and vacancy-bearing (220) lattice planes (highlighted). Both structural types accommodate trigonal pyramids of Fe-atoms as a common building block, forming a pyrochlore-type sublattice in magnetite.

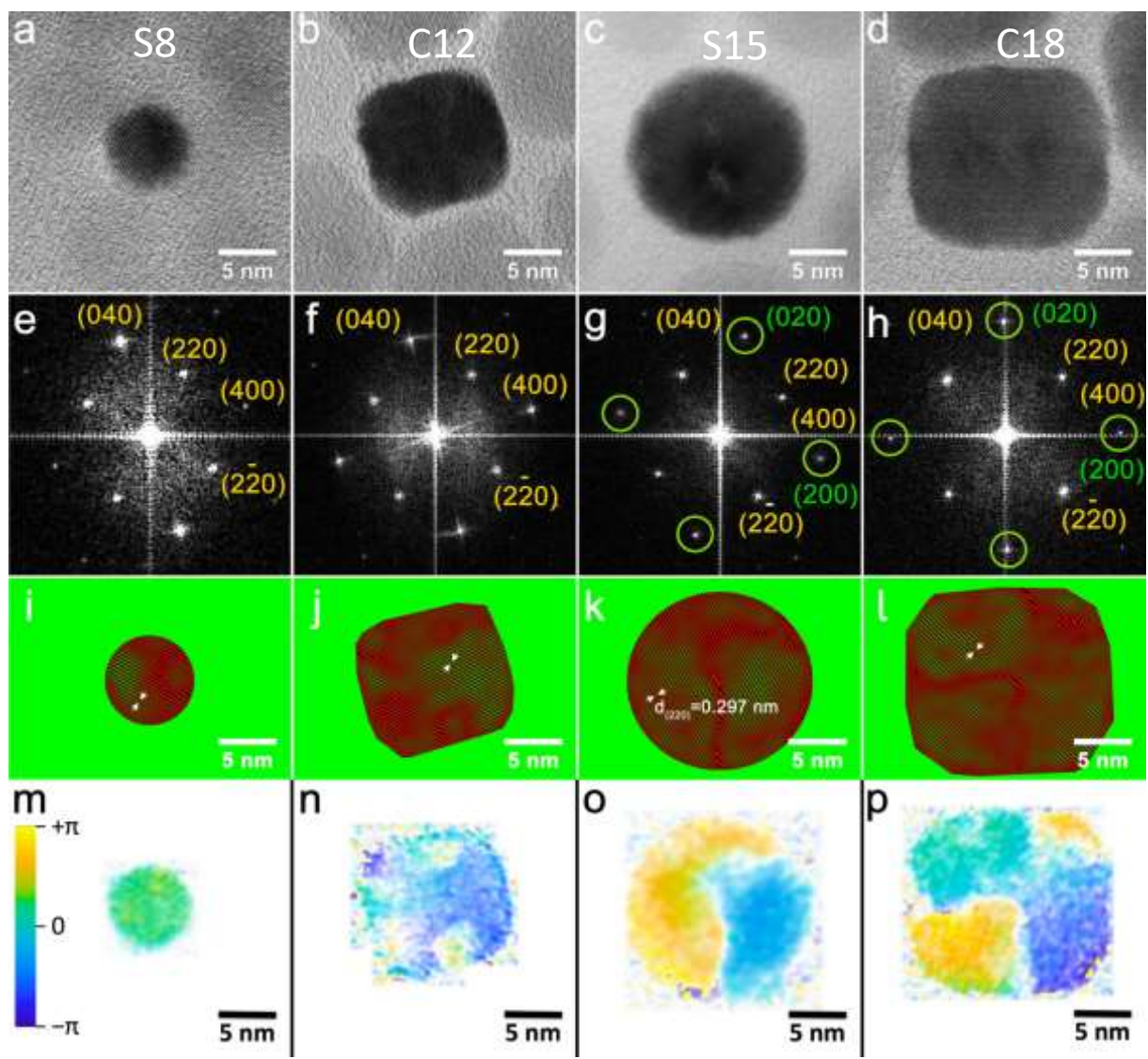

**Figure 2.** High-resolution TEM images in the [001] zone axes for spherical S8 (a), S15 (c) and cubic C12 (b), C18 (d) morphology nanocrystals. The corresponding diffraction patterns (e, f and h, g) after Fast Fourier Transform (FFT) analysis of each micrograph are shown beneath. Green and yellow circles mark set of reflections that could be indexed on the basis of either wüstite (green) or/and magnetite (yellow) rock-salt and cubic spinel crystal unit cells (see text for details). Representative real space images of the (220) atomic lattice planes (inverse FFT synthesis; filtered from specific plane orientations) for the samples S8 (i), C12 (j), S15 (k) and C18 (l), respectively. Lattice planes have been colored with red (possible presence of atomic plane) and green (possible absence of atomic plane) pseudo-chrome acquired after inverse FFT process. Lattice phase contrast images obtained by the GPA

method (see text) after re-centering the diffraction around one of the (220) reflections for samples S8 (m), C12 (n), S15 (o) and C18 (p).

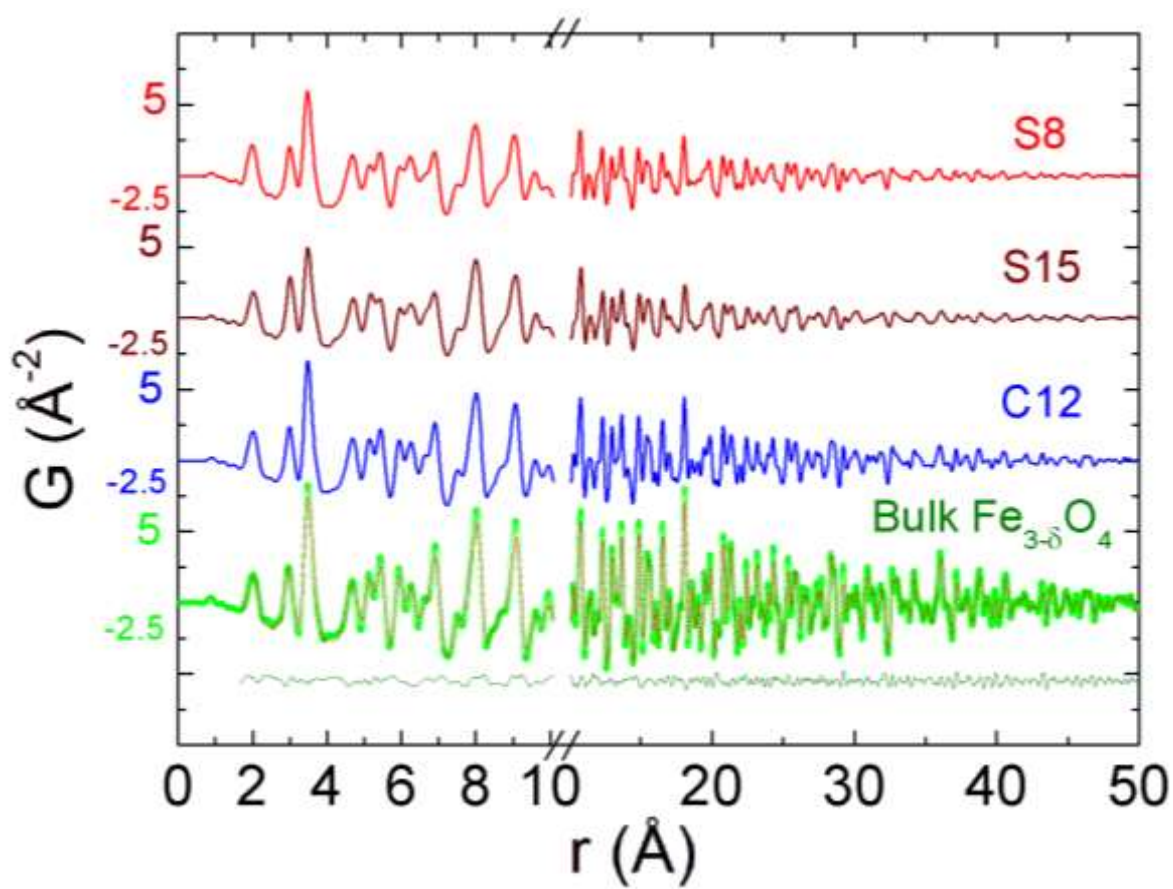

**Figure 3.** Experimental atomic xPDF data at 80 K for the single-phase S8, C12 and core@shell S15 nanoparticles plotted as a function of the radial distance, r up to 50 Å. The corresponding xPDF pattern for the bulk magnetite, measured under the same exact conditions, is provided as a reference material. The line over the data is the best fit based on the cubic spinel atomic configuration model (Fd-3m symmetry, model #1 and $R_w$= 11.6 %, the quality of fit factor) and below the corresponding difference between observed and calculated xPDFs for bulk magnetite.

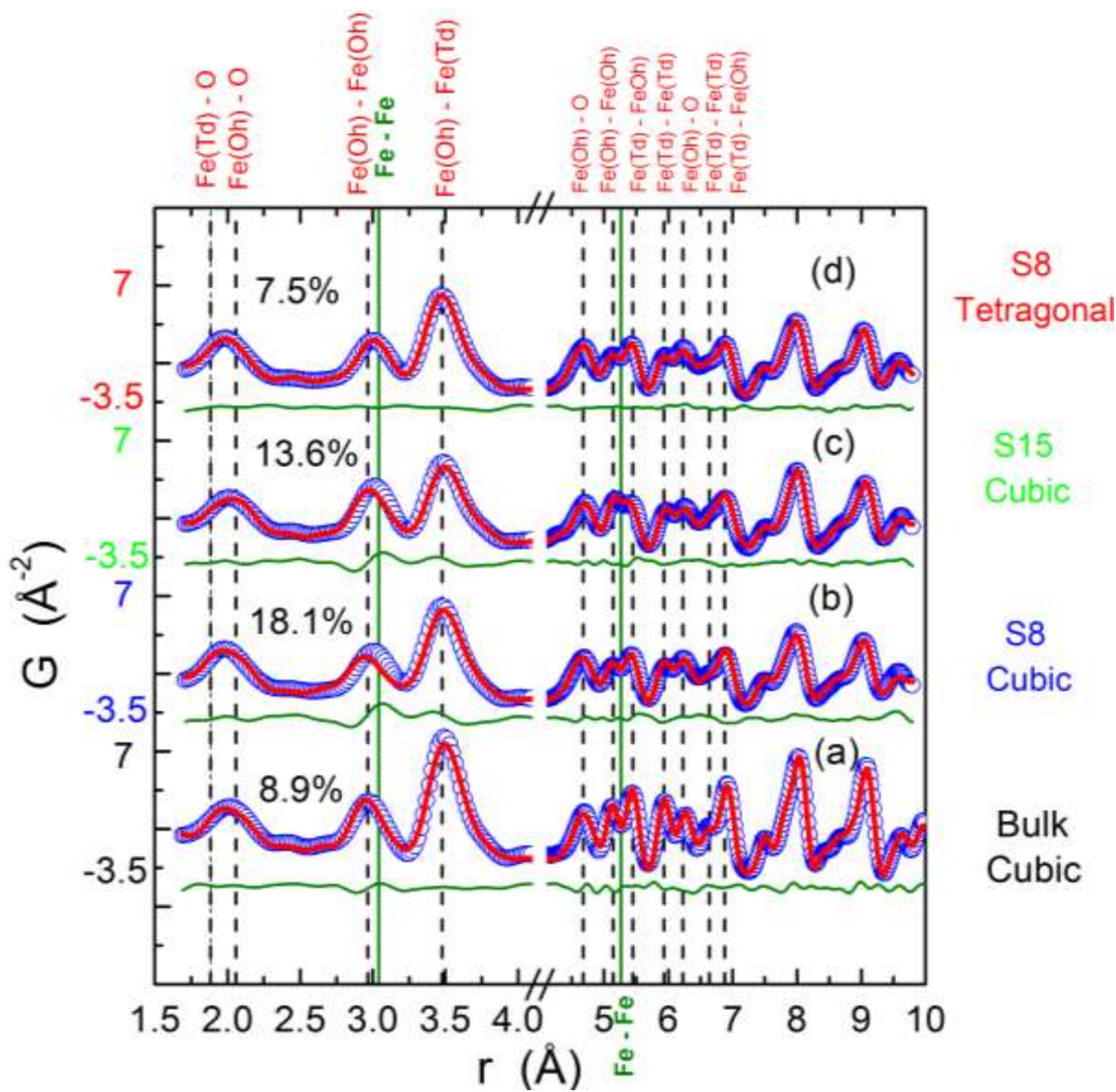

**Figure 4.** Representative xPDF fits (T= 80 K) of the "low-r" region for (a) the bulk magnetite assuming the cubic spinel atomic configuration (Fd-3m, model #1; a= 8.3913(2) Å, $R_w$= 8.9 %), (b) the single-phase S8 nanoparticle sample with the same cubic model ($R_w$= 18.1 %), (c) the two-phase S15 nanoparticle sample with the cubic rock-salt and spinel ($R_w$= 13.6 %) crystallographic models, and (d) the single-phase S8 nanoparticle sample in the tetragonal (P4$_3$2$_1$2, model #3; $a_{sp}$=$b_{sp}$=8.4053(1) Å, $c_{sp}$= 8.1950(1) Å, $R_w$= 7.5%) crystallographic configuration (see text for details). The blue circles and red solid lines correspond to the observed and simulated atomic PDFs respectively. The green solid lines underneath are the difference curves between observed and calculated PDFs. The quality of fit factor, $R_w$ (%), is given for each case. Vertical dashed lines mark the positions of typical Fe-O and Fe-Fe bond distances; Td and Oh represent tetrahedral and octahedral cation sites, respectively.

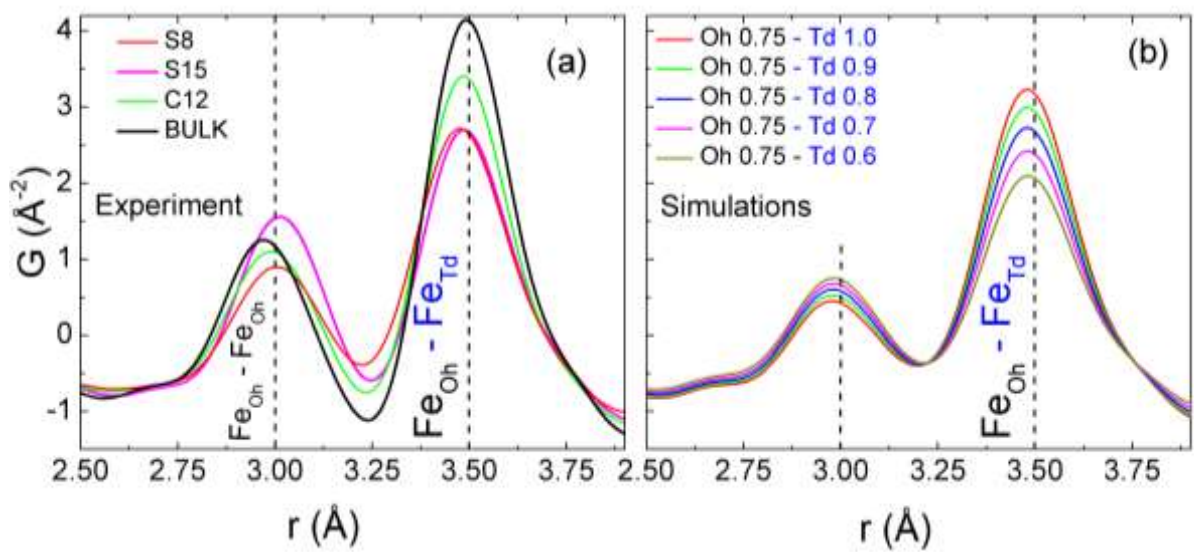

**Figure 5.** (a) Adequately normalised, observed G(r) patterns (T= 80 K) of the low-r region for all nanoparticle samples (single-phase, S8 & C12 and two-phase, S15) compared against the bulk magnetite and (b) the simulated xPDF patterns on the basis of the symmetry-lowering (tetragonal) atomic configuration. As a proof of concept, the optimally chosen set of models assumes a sub-stoichiometric Oh Fe-site occupancy ($\eta$; kept constant at 75%, an average occupation derived from refinements of the G(r) data across the samples), while the Td Fe-site occupation level was varied in a step-wise manner (60<$\eta$<100 %). Vertical dashed lines mark the modification of the distribution of the $Fe_{Oh}$-$Fe_{Oh}$ ($r \sim 3.0$ Å) and $Fe_{Oh}$-$Fe_{Td}$ ($r \sim 3.5$ Å) separations from bulk to nanoscale samples.

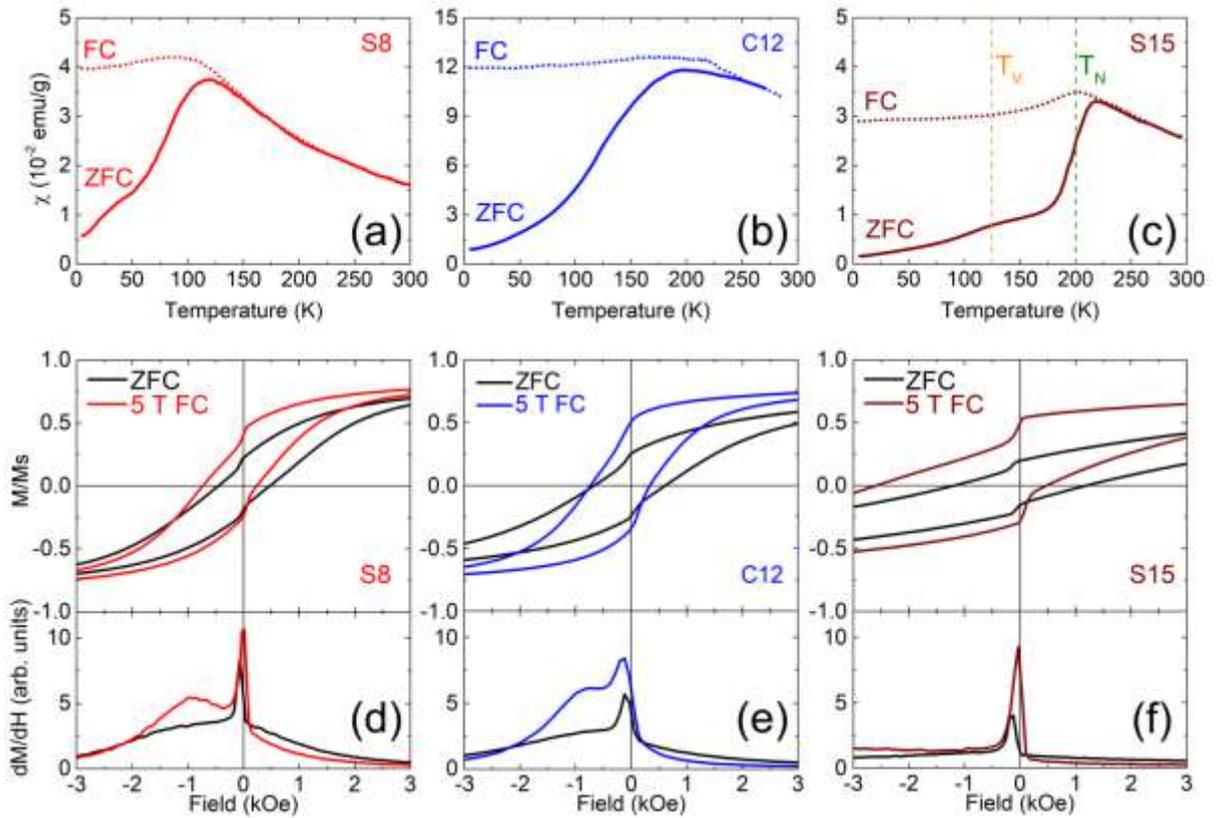

**Figure 6.** The temperature evolution of the zero-field cooled (ZFC; solid lines) and field-cooled (FC; dotted lines) susceptibility curves for the single-phase S8 (a), C12 (b) and core@shell S15 (c) nanoparticles under a magnetic field of 50 Oe. The dashed lines indicate the Verwey ($T_V$; orange) and Néel ($T_N$; green) related transition temperatures met in bulk, stoichiometric $Fe_3O_4$ and $Fe_xO$, respectively. Low-field part of the normalized hysteresis loops ($M/M_s$) at 5 K for the single-phase S8 (d), C12 (e) and the core@shell S15 (f) nanoparticles taken under zero- and field- cooled ($H_{cool}$= 5 T) protocols. The panels beneath the loops present the differential change ($dM/dH$) of the normalized, isothermal magnetization when it is switched from positive to negative saturation.

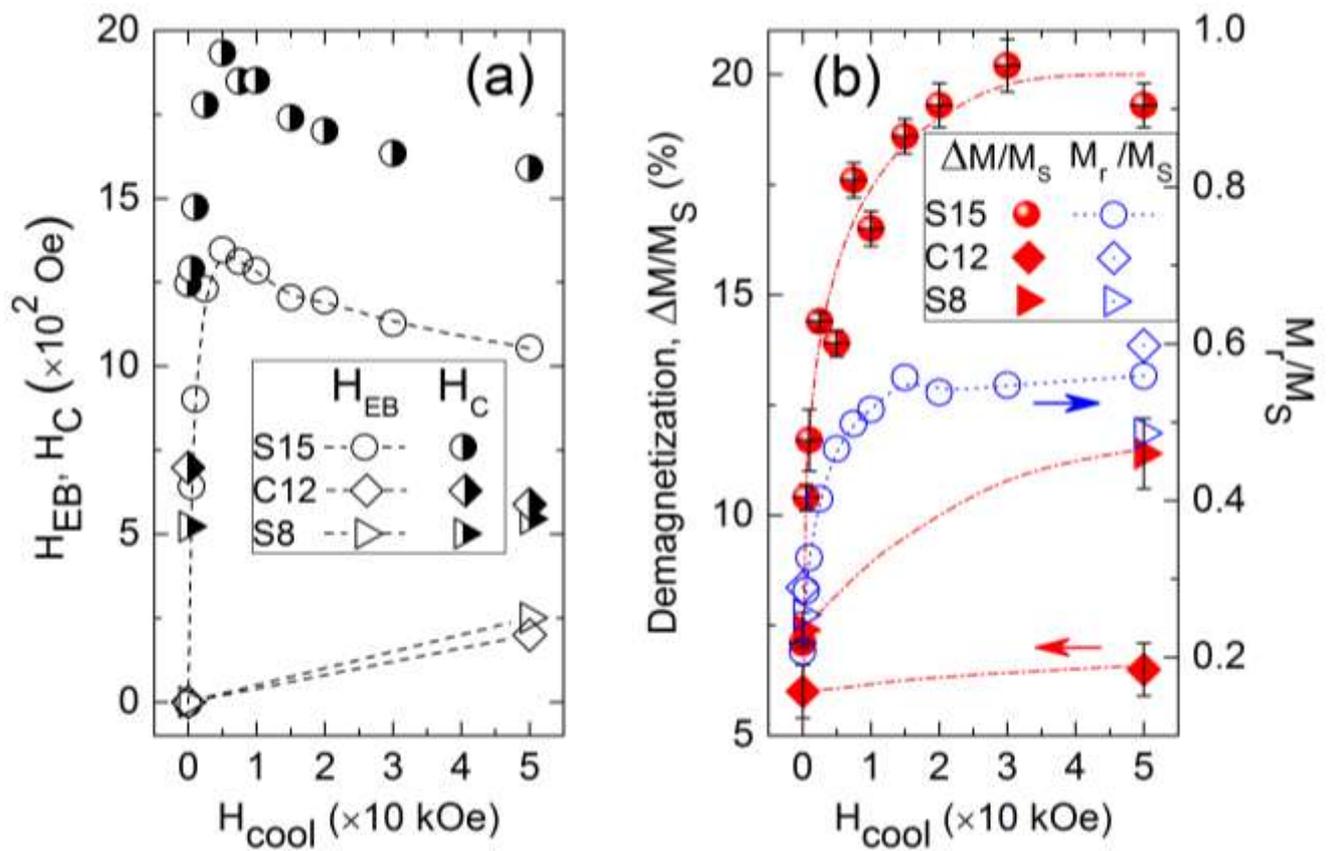

**Figure 7.** The experimentally determined (a) exchange-bias ($H_{EB}$; open symbols) and coercive field ($H_C$; half-filled symbols), as well as (b) low-field demagnetization ($\Delta M/M_S$; filled symbols, left y-axis) and the ratio $M_r/M_S$ ($M_r$, remanence: open symbols, right y-axis) obtained at varying cooling-field strengths ($H_{cool}$). Symbol guide: circular, core-shell spherical nanoparticles (S15), triangular (S8) and diamond (C12) for single-phase spherical and cubic morphology nanoparticles. The lines are a guide to the eye.

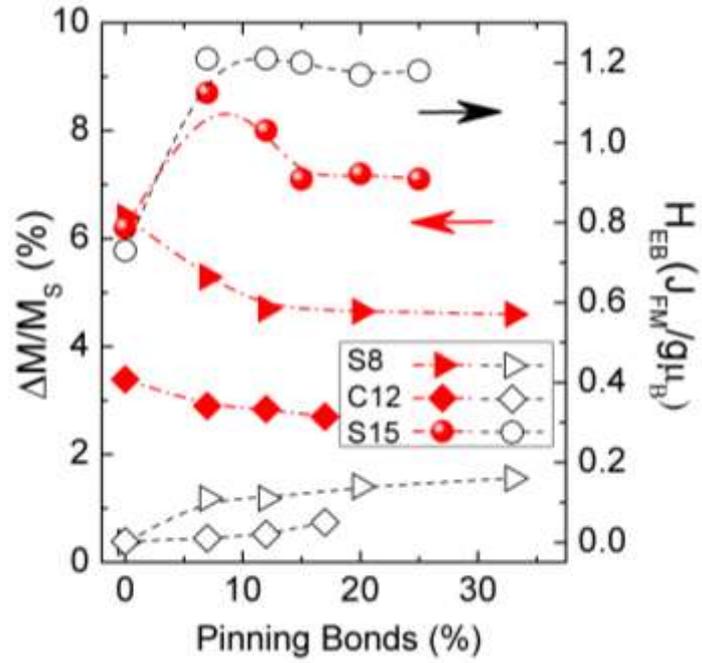

**Figure 8.** Monte Carlo calculations ($H_{cool}$= 5 $J_{FM}/g\mu_B$) of the effect of pinning bonds (vacancy-driven) on the low-field magnetic moment switching ($\Delta M/M_S$; filled symbols, left y-axis) and the exchange-bias ($H_{EB}$; open symbols, right y-axis) of self-passivated $Fe_xO$-$Fe_3O_4$ nanocrystals with different morphologies. Symbol guide: circular, core-shell spherical nanoparticles (S15), triangular (S8) and diamond (C12) for single-phase spherical and cubic shape nanoparticles.

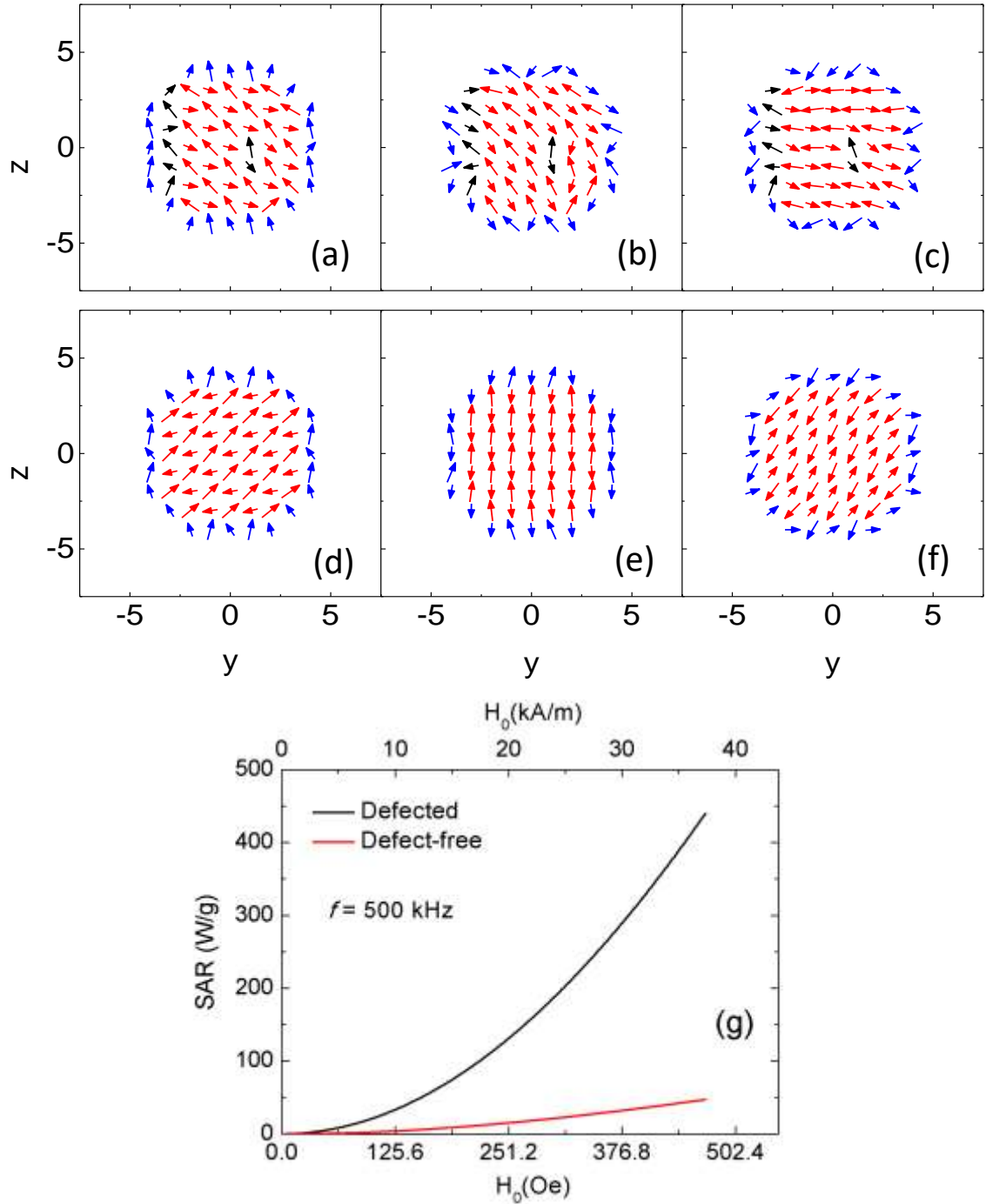

**Figure 9.** Monte Carlo (MC) simulation of the *M-H* loops made after field-cooling ($H_{cool}$= 5 $J_{FM}$/g$\mu_B$), and by assuming an upper maximum fraction of pinning bonds, as this was identified by the xPDF refinements. Snap-shots of the spin-ensemble during *M-H* loop calculation for a small ferrimagnetic spherical nanoparticle of R= 5 lattice constants: (a-c) a fully oxidized, defected nanocrystal (S8), and (d-f) a defect-free case, assuming a core-surface type of MC model. Spin configurations are shown in the zy-plane (x= −1) for positive field (*H*= +8 $J_{FM}$/g$\mu_B$) magnetization saturation and after field reversal (*H*= − 0.2 $J_{FM}$/g$\mu_B$, *H*= − 2.0 $J_{FM}$/g$\mu_B$) towards negative saturation. The arrow color-coding stands for: core (red), surface (blue), defects (black) magnetic moments. (g) Specific Absorption Rate (SAR) of small defected (S8) versus defect-free nanoparticles on the basis of susceptibility losses (AC field amplitude, $H_0$ and *f*= 500 kHz), calculated according to the Linear Response Theory of a modified Néel-Brown relaxation Monte Carlo model.

# SUPPLEMENTAL MATERIAL

**Vacancy-driven non-cubic local structure and magnetic anisotropy tailoring in $Fe_xO$-$Fe_{3-\delta}O_4$ nanocrystals**

Alexandros Lappas,[1,*] George Antonaropoulos,[1,2] Konstantinos Brintakis,[1] Marianna Vasilakaki,[3] Kalliopi N. Trohidou,[3] Vincenzo Iannotti,[4] Giovanni Ausanio,[4] Athanasia Kostopoulou,[1] Milinda Abeykoon,[5] Ian K. Robinson[6,7] and Emil S. Bozin[6]

[1]Institute of Electronic Structure and Laser, Foundation for Research and Technology - Hellas, Vassilika Vouton, 71110 Heraklion, Greece

[2]Department of Chemistry, University of Crete, Voutes, 71003 Heraklion, Greece

[3]Institute of Nanoscience and Nanotechnology, National Center for Scientific Research Demokritos, 15310 Athens, Greece

[4]CNR-SPIN and Department of Physics "E. Pancini", University of Naples Federico II, Piazzale V. Tecchio 80, 80125 Naples, Italy

[5]Photon Sciences Division, National Synchrotron Light Source II, Brookhaven National Laboratory, Upton, 11973 NY, USA

[6]Condensed Matter Physics and Materials Science Department, Brookhaven National Laboratory, Upton, 11973 NY, USA

[7]London Centre for Nanotechnology, University College, London WC1E 6BT, UK

[*]Correspondence: lappas@iesl.forth.gr; Tel.: +30 2810 391344

# Table of Contents

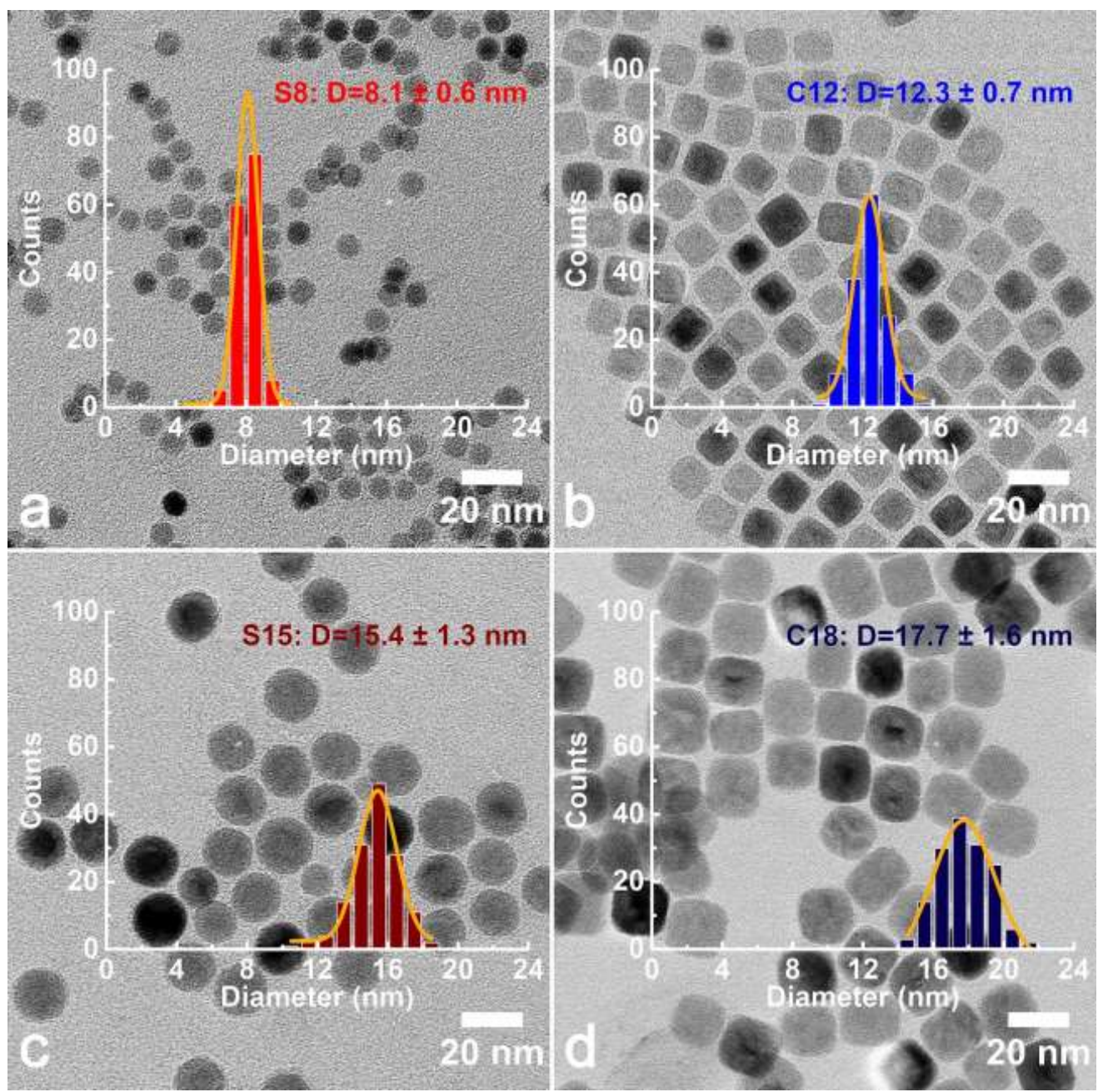

**Figure S1.** Low-magnification bright-field transmission electron microscopy (TEM) images of iron-oxide nanocrystals of spherical (a) S8, (c) S15 and cubic (b) C12, (d) C18 morphologies. The nanoparticle size distributions are shown as insets to the micrographs.

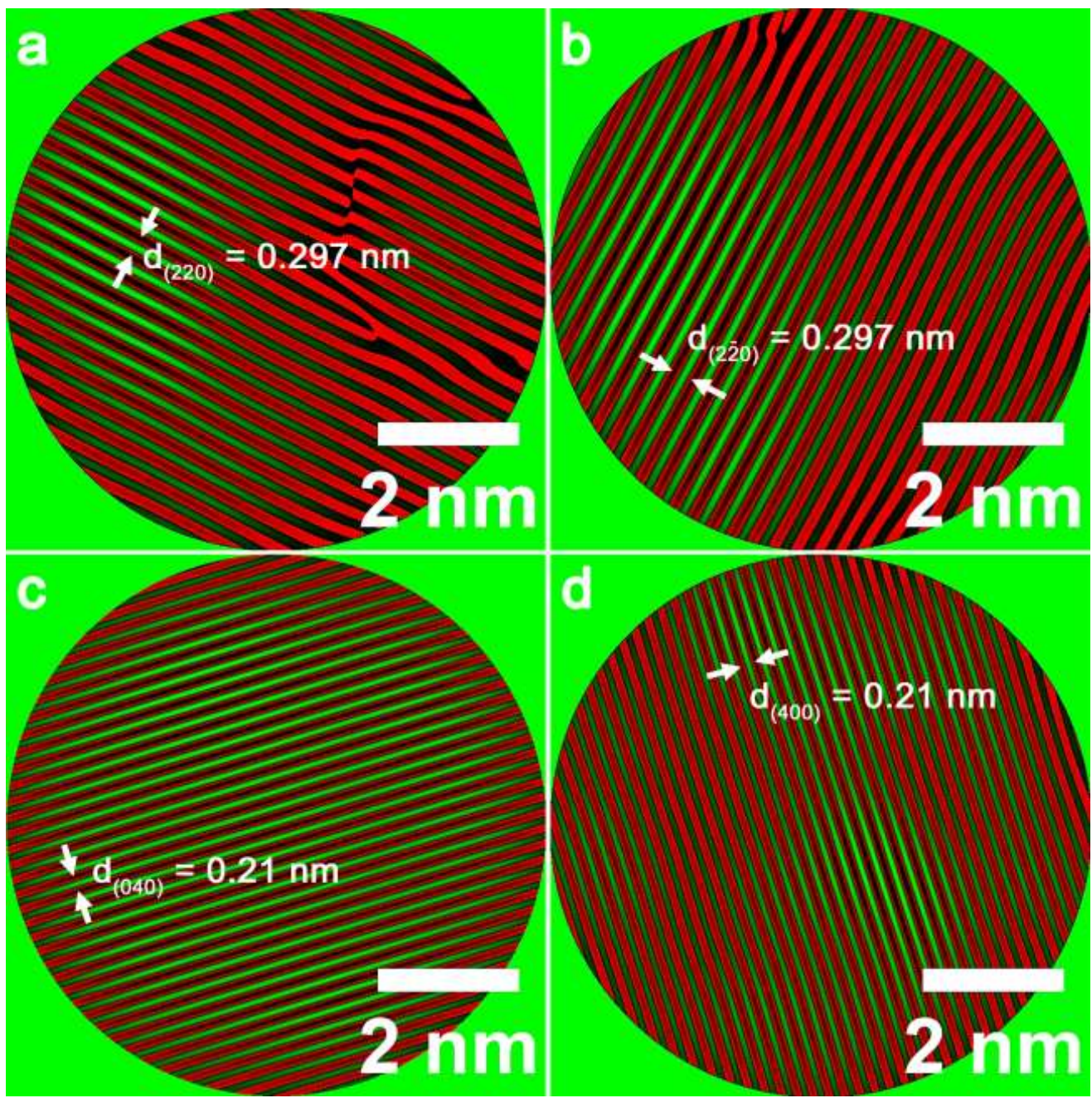

**Figure S2.1.** Representative real space images of S8 single-phase spherical nanoparticle, with diameter 8.1 nm. Lattice planes colored with red, green pseudo-chrome after inverse FFT synthesis of the diffraction pattern shown in Fig. 1e. (a) (220), (b) (2-20), (c) (040) and (d) (400)atomic lattice planes in the [001] zone axes.

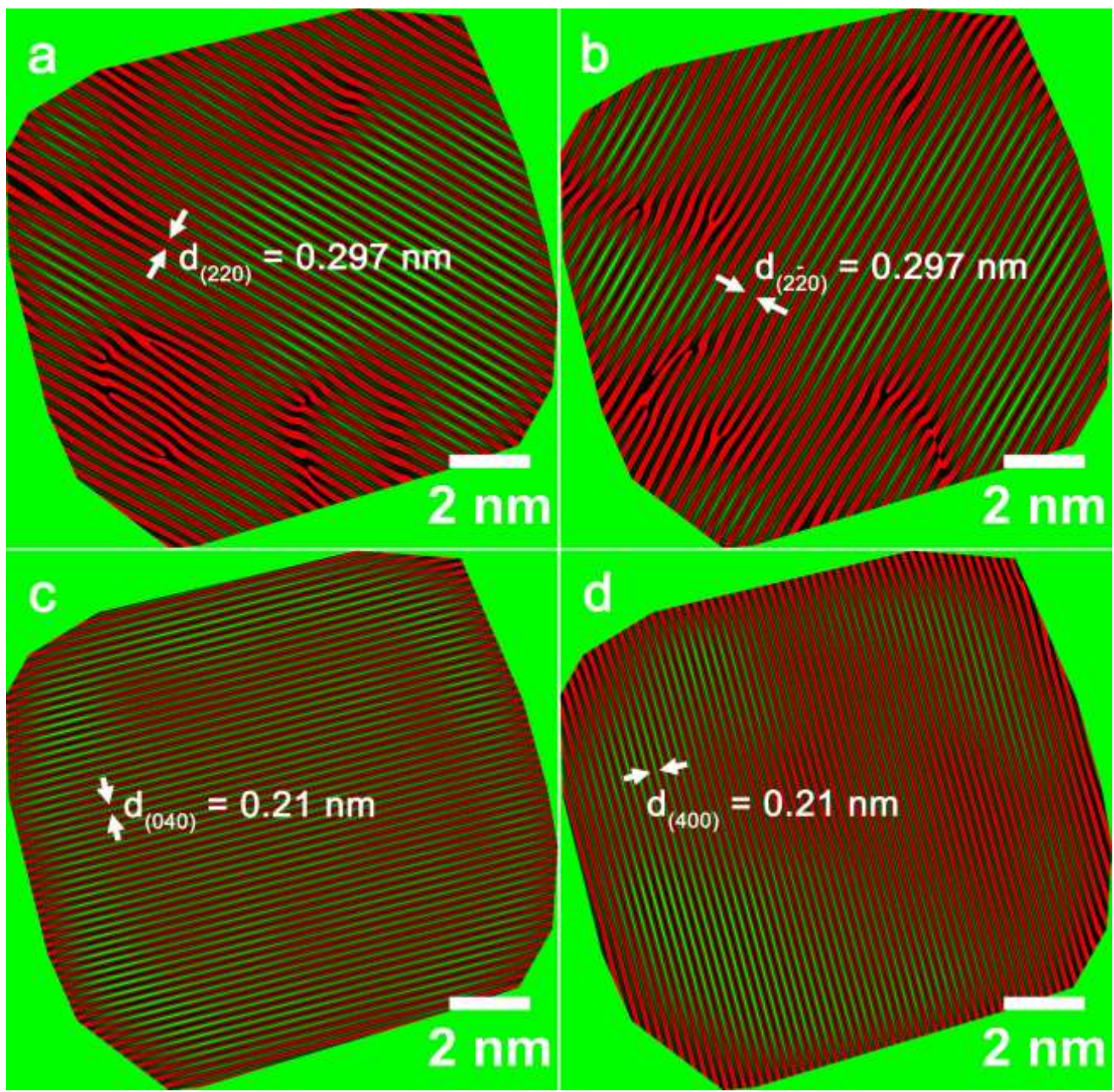

**Figure S2.2.** Representative real space images of C12 single phase cubic $Fe_3O_4$ nanoparticle with edge length 12.3 nm. Lattice planes colored with red, green pseudochrome after inverse FFT of the image shown in Fig. 1f. (a) (220), (b) (2-20), (c) (040) and (d) (400) lattice planes in zone axes [001].

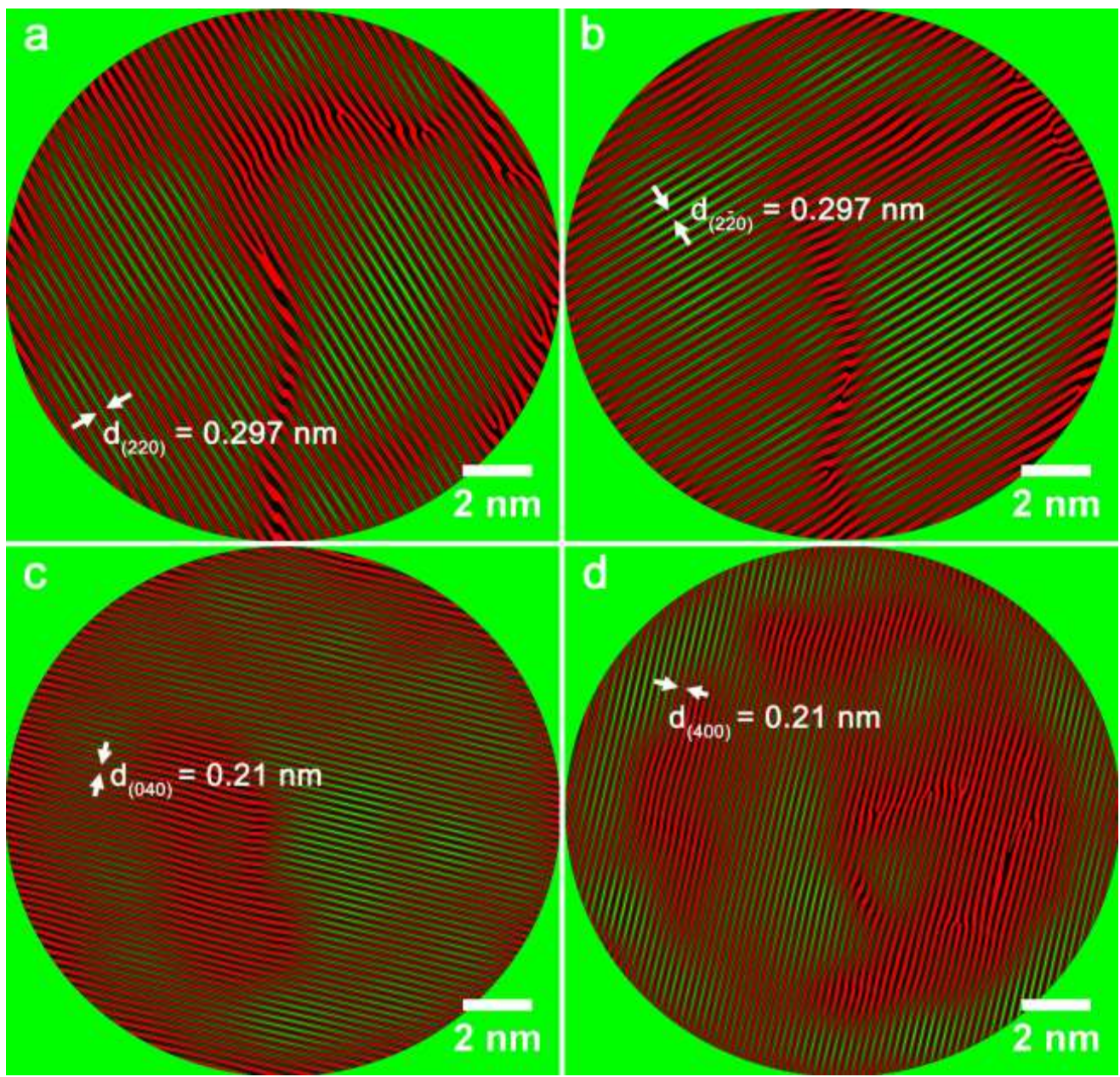

**Figure S2.3.** Representative real space images of S15 spherical core@shell FeO@Fe$_3$O$_4$ nanoparticle with diameter 15.4 nm. Lattice planes colored with red, green pseudochrome after inverse FFT of the image shown in Fig. 1g. (a) (220), (b) (2-20), (c) (040) and (d) (400) lattice planes in zone axes [001]; in panels (c) and (d) lattice fringes can be attributed also to d$_{(020)}$ = d$_{(200)}$ = 0.21 nm of the rock-salt FeO phase.

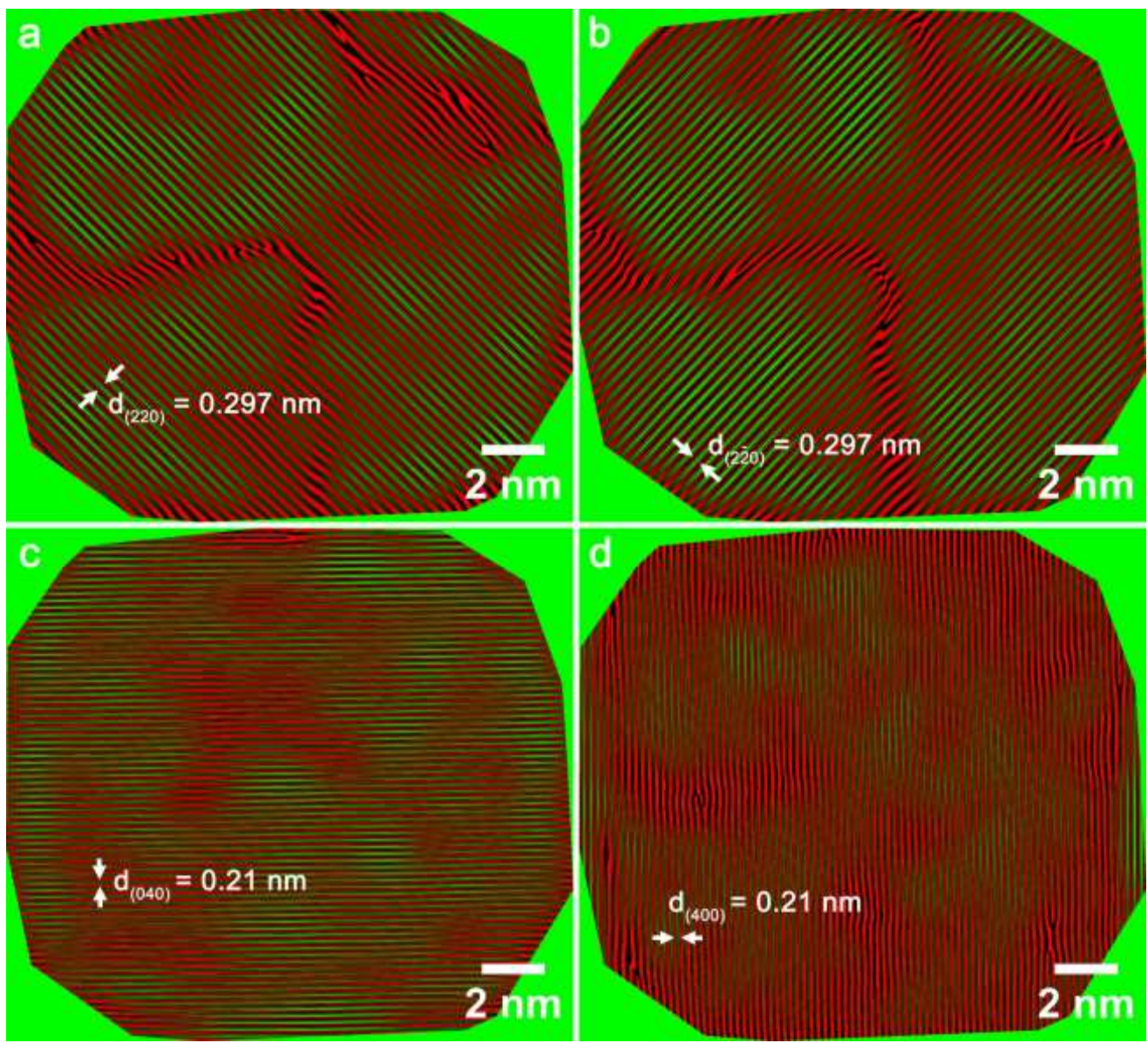

**Figure S2.4.** Representative real space images of C18 cubic core@shell FeO@Fe$_3$O$_4$ nanoparticle with diameter 17.7 nm. Lattice planes colored with red, green pseudochrome after inverse FFT of the image shown in Fig. 1h. (a) (220), (b) (2-20), (c) (040) and (d) (400) lattice planes in zone axes [001]; in panels (c) and (d) the lattice fringes can also be attributed to $d_{(020)} = d_{(200)} = 0.21$ nm of the rock-salt FeO phase.

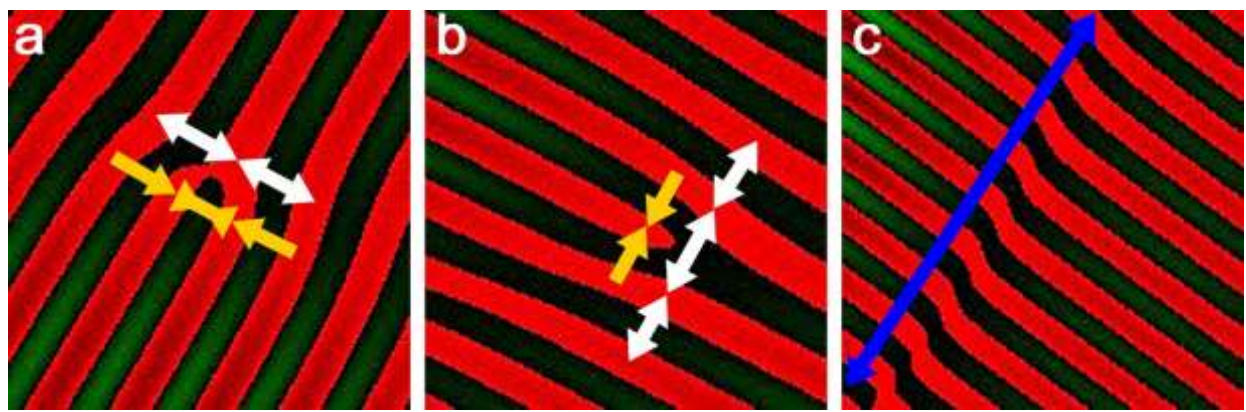

**Figure S3.** (a) Screw, (b) edge, (c) slip dislocations. Yellow arrows depict lattice contraction, white arrows lattice expansion and blue one the slip length.

## S3. HRTEM – inverse FFT synthesis – No. of defects

The findings that the (220) atomic planes suffer from three types of defects (Fig. S3) motivated a subsequent statistical analysis. This attempted to account for the presence of the number (Nr) of defects involving the (220) planes, throughout the volume of the nanoparticles with different size and shape. Due to the tedious nature of this analysis only a relatively small number of 8-10 NPs has been counted upon, as compared to the much larger ensemble (100-150 NPs) utilized to extract the TEM particle size distributions (Fig. S1). Table S1 and Figure S4 summarize the outcome of the analysis. Despite the limited statistics involved, cautious interpretation points that NPs of spherical morphology (S8, S15) appear to carry a larger number of defects in their volume, and those of smaller size (S8) are somewhat more susceptible to lattice faults.

| Sample Label | Avg Size (nm) | $\sigma_{size}$ (nm) | Volume (nm³) | Nr of counted NPs | Total measured Nr of Defects | Nr Def/ NP Volume (%) | Error (%) |
|---|---|---|---|---|---|---|---|
| S8 | 8.4 | 0.5 | 317 | 10 | 88 | 2.8 | 1.4 |
| C12 | 13.7 | 0.8 | 2573 | 10 | 180 | 0.7 | 0.4 |
| S15 | 16.7 | 0.9 | 2438 | 9 | 366 | 1.6 | 0.8 |
| C18 | 18.7 | 1.2 | 6586 | 8 | 411 | 0.8 | 0.3 |

**Table S1.** Calculated parameters for defects. Average size of NPs, volume of the NPs, number (Nr) of accounted nanoparticles, total Nr of defects of the two (220) lattice planes of $Fe_3O_4$ and Nr of defects per NP's volume. Reddish rows show the spherical NPs and bluish the cubic ones.

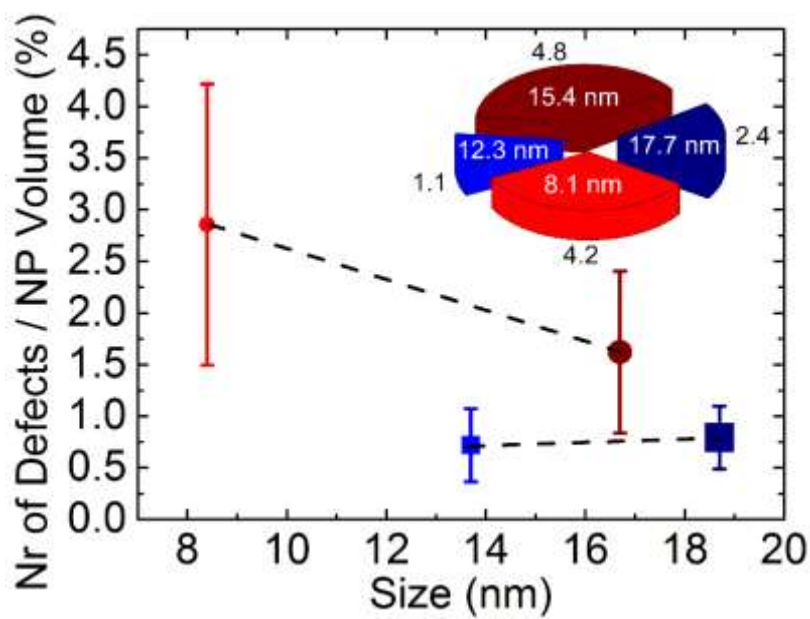

**Figure S4.** Size-dependence of the total number of nanoparticle defects normalized by their volume. Reddish circles depict the spherical nanocrystals, while the bluish squares the cubic ones. Inset: Pie-chart depicts the tensile strain (%) in the total volume of each type of nanoparticle. Reddish sections correspond to the spherical nanocrystals and bluish to those of cubic morphology.

### S4. HRTEM – inverse FFT synthesis – strain

The afore-mentioned, distorted (220) atomic planes are likely to impose non-negligible lattice strains. In order to obtain a semi-quantitative description of such a distortion, we estimated the lattice modification near the screw and edge dislocations relative to the undistorted spacing of $d_{(220)Fe3O4}$ = 2.97 Å that exists around each type of defect. Then it is possible to arrive to an assessment of how extensively this distortion populates each type of nanoparticle morphology by normalizing the obtained values with the number of defects per particle volume. In this approximation the type of lattice strain, is identified by the sign of the relative lattice distortion (Table S2), which can be either positive for tensile or negative for compressive modifications. The degree of lattice strain for NPs with different sizes and shapes is depicted in Figure S4 (inset). Although the screw dislocations appear to cause tensile strain around the defected area and the edge dislocations generate compressive strain, in all variety of particles made here, tensile strain in the (220) lattice planes seems to be the overall prevailing effect that is also pronounced in spherical morphology.

As depicted in Figure S3, both expansion and contraction are observed around each screw and edge dislocation. We define the strain as the difference between the two types of distortions. Expansion is assumed when the lattice planes are separated more than the $d_{(220)Fe3O4} = 2.97$ Å of the bulk and contraction when they are nearer than that reference value. Tensile or compressive strain is marked by the sign of difference and is compiled in Table S2, for the Screw (3rd column) and Edge (6th column) dislocations as Disl. Exp.-Cont.. The Slips are recorded with respect to the according the nanoparticle "length" of the (10th column). In order to see the effect of strain in the total volume of the nanoparticle, the values were normalized (columns 5th, 8th, their sum 9th and 12th for the slips) by the number of defects per NP's volume.

| 1 | 2 | 3 | 4 | 5 | 6 | 7 | 8 | 9 | 10 | 11 | 12 |
|---|---|---|---|---|---|---|---|---|----|----|----|
| Sample Label | Avg Size (nm) | Screw Disl. Exp.-Cont. (%) | Nr Screw Disl. per Volume (%) | Norm. Screw Disl. (%) | Edge Disl. Exp.-Cont. (%) | Nr Edge Disl. per Volume (%) | Norm. Edge Disl. (%) | Norm. Sum Screw & Edge (%) | Slip Length per Length of NP (%) | Nr Slip Disl. per Volume (%) | Norm. Slip Disl. (%) |
| S8 | 8.4 | 8.4 | 0.86 | 7.2 | -2.9 | 1.03 | -3.0 | 4.2 | 19.6 | 0.96 | 18.8 |
| C12 | 13.7 | 6.5 | 0.25 | 1.6 | -3.0 | 0.16 | -0.5 | 1.1 | 11.4 | 0.31 | 3.5 |
| S15 | 16.7 | 11.1 | 0.47 | 5.2 | -0.7 | 0.57 | -0.4 | 4.8 | 10.1 | 0.57 | 5.8 |
| C18 | 18.7 | 9.4 | 0.27 | 2.5 | -0.6 | 0.30 | -0.2 | 2.4 | 8.7 | 0.23 | 2.0 |

**Table S2.** Percentages of sum (expansions – contractions) of (220) lattice planes near each dislocation (screw & edge disl.) normalized to the Nr of defects per volume of NPs. Percentage of slip dislocations' length per length or diameter of a NP for (220) lattice planes normalized to the Nr of slips per volume of NPs. Reddish rows show the spherical NPs and bluish the cubic ones.

## S5. Atomic PDF analysis

The data of all NP samples and the bulk $Fe_{3-\delta}O_4$ reference were initially modeled on the basis of the normal cubic spinel structure (model #1; Table S3). [1] The iron atoms were placed at the crystallographic 8$a$ (Td-site) and 16$d$ (Oh-site) positions in the space group Fd-3m, while the oxygen occupied the 32$e$ site, with initial coordinates (0.255, 0.255, 0.255). A second phase was included for the larger core@shell particles to account for the FeO phase, with iron at 4$a$ (0, 0, 0) and oxygen at 4$b$ (1/2, 1/2, 1/2) sites. Lattice parameters alone were refined initially in the model. Atomic site occupancies were set to reflect the sample composition in the ideal crystal structure. The overall scale factor was refined in each case, with additional scale factor added to account for the phase fraction in two-phase models. Other parameters which affect the quality of the fit and are related to Q-space resolution effects, such as PDF signal-dampening and peak-broadening were accounted for. These parameters, related to the experimental setup, were determined from a Ni standard sample measurement and kept fixed. Sample dependent parameters were refined appropriately in all models. Subsequently, all the atomic positions released, followed by the thermal factors. The thermal vibrations were assumed to be isotropic. Different parameters for the thermal factors were assigned to the crystallographically distinct Fe- sites. Atom site occupancies for iron in the spinel crystal lattice were the last parameters to be refined, and were set to be independent for distinct crystallographic positions. Site occupancies for oxygen were not refined due to lack of relative X-ray sensitivity for such lighter atoms. Thermal factors and site occupancies were found to be strongly correlated; thus, they were not refined simultaneously. In order to calculate the Fe-site occupancies and to assure reliable comparisons, the thermal factors were fixed to values determined from the $Fe_{3-\delta}O_4$ bulk reference; these were set to be identical for all samples studied, while the occupancies were freely refined.

As discussed in the main body of the manuscript (§1.2.1), this initial model proved to be insufficient in describing the local structure (r= 1-10 Å) of the NP systems studied. Model #3, based on the P4$_3$2$_1$2 tetragonal symmetry (Table S3), [2] made a significant improvement in the studied r-range and thus it was utilized for the presented quantifications. The fitting procedure explained above for model #1, was then attended for the symmetry-lowering model #3. This tetragonal model assumes four distinct crystallographic positions for iron: one at a modified Td-site and three different Oh-sites (Table S3). The initial atomic coordinates of model #3 and their PDF-refined values for a locally distorted NP are compiled in Table S3. For ease of comparison the normal cubic spinel, initial crystallographic parameters for the bulk $Fe_{3-\delta}O_4$ sample and their PDF-refined counterparts are also included in Table S3. Additional results from fitting the low-r region atomic PDF (r= 1-10 Å) for all samples are summarized in Table S4.

The PDF analysis suggest that the local structure, in the low-r region, is better described by the tetragonal model (model #3), as indicated by the lower fit residuals ($R_w$). Since the presented analysis is limited to sub-nanometer length-scale only, it should be noted that this does not address nor suggest the existence of a distinct, ideally structured tetragonal phase in the NP samples. The tetragonal model was utilized as a proxy to assess the very local structural aspects and facilitate the quantifications presented. In order to determine if the tetragonal model provides an adequate description of the NP structure as a whole, we evaluated how well the tetragonal model performs over different length-scales, compared to the cubic one. This was done by conducting the so-called "box-car" PDF-refinement tests. In the latter, the experimental G(r) data are analyzed in different, sequential r-windows, using the same crystallographic model and fitting parameters. This allows for an assessment of the consistency of the chosen model over variable length-scales. The results of box-car refinement fit-residuals ($R_w$) for the nanoparticle sample S8 are presented in Fig. S5.

These indicate that for the S8 there is a localized, short-range distortion, induced by sub-stoichiometry (Fe-vacancy), which is better described in the r-range 1-10 Å by a more parameterized model, such as model #3 (Fig. S5). The moderate Q-space resolution of the experimental setup limits the PDF field of view and leads to progressively less reliable results in the r-space above 10-11 Å. In fact, none of our models (cubic *vs.* tetragonal) could model the crystal structure of our nanoparticle samples at radial distances longer than 10 Å. We suspect that there are strains associated with the defects (§1.1), which are impossible to model using the small-box modeling methodology utilized here. These could perhaps be addressed using approaches that are beyond the scope of this study. Nevertheless, the results of the box-car type of refinement (Fig. S5) indicate that the locally distorted structure possibly relaxes to the expected cubic spinel structure in the high-r region, suggesting that the average structure of the samples is "cubic-like" at high-r. Even though the quality of the fit is far from optimal in this region, the cubic spinel model (model #1) appears to be far more consistent than the tetragonal one (model #3).

| TETRAGONAL (P4$_3$2$_1$2) | | | | | | | CUBIC (Fd-3m) | | |
|---|---|---|---|---|---|---|---|---|---|
| | **S8** | | | **Model #3** [2] | | | | **Bulk $Fe_{3-\delta}O_4$** | **Model #1** [1] |
| **a (Å)** | 8.4053(1) | | | 8.3396 | | | **a (Å)** | | |
| **b (Å)** | | | | | | | **b (Å)** | 8.3913(2) | 8.395 |
| **c (Å)** | 8.1950(1) | | | 8.322 | | | **c (Å)** | | |
| **V (Å$^3$)** | 578.9691(1) | | | 578.7862 | | | **V (Å$^3$)** | 590.8643(1) | 591.6462 |
| **Fe-Td** | | | | | | | **Fe-Td** | | |
| **8$b$** | 0.741(2) | 0.995(1) | 0.122(1) | 0.744 | 0.996 | 0.12 | **8$a$** x=y=z | 0.125 | 0.125 |
| **Uiso (Å$^2$)** | 0.0030(1) | | | 0.001 | | | **Uiso (Å$^2$)** | 0.0041(1) | 0.0063 |
| **Fe atoms/ unit cell** | 6.30(4) | | | 8 | | | **Fe atoms/ unit cell** | 9.40(1) | 8 |
| **Fe-Oh$_1$** | | | | | | | **Fe-Oh** | | |
| **4$a$** | 0.624(1) | 0.624(1) | 0 | 0.62 | 0.62 | 0 | | | |
| **Uiso (Å$^2$)** | 0.0029(3) | | | 0.001 | | | | | |
| **Fe-Oh$_2$** | | | | | | | **16$d$** x=y=z | 0.5 | 0.5 |
| **8$b$** | 0.368(2) | 0.874(1) | 0.983(2) | 0.364 | 0.867 | 0.984 | | | |
| **Uiso (Å$^2$)** | 0.0032(1) | | | 0.001 | | | | | |
| **Fe-Oh$_3$** | | | | | | | | | |
| **4$a$** | 0.164(2) | 0.164(2) | 0 | 0.14 | 0.14 | 0 | **Uiso (Å$^2$)** | 0.0052(1) | 0.0081 |
| **Uiso (Å$^2$)** | 0.0014(2) | | | 0.001 | | | | | |
| **Fe atoms/ unit cell** | 10.01(3) | | | 16 | | | **Fe atoms/ unit cell** | 17.49(1) | 16 |
| **Total Fe atoms/ unit cell** | 16.31(6) | | | 24 | | | **Total Fe atoms/ unit cell** | 26.891(1) | 24 |
| **O$_1$**   **8$b$** | 0.604(2) | 0.856(4) | 0.989(3) | 0.615 | 0.869 | 0.986 | | | |
| **O$_2$**   **8$b$** | 0.109(5) | 0.363(1) | 0.993(3) | 0.119 | 0.377 | 0.995 | **O**   **32$e$** x=y=z | 0.2552(1) | 0.2546 |
| **O$_3$**   **8$b$** | 0.129(3) | 0.867(3) | 0.999(3) | 0.137 | 0.861 | 0.007 | | | |
| **O$_4$**   **8$b$** | 0.361(2) | 0.624(7) | 0.999(5) | 0.383 | 0.631 | 0.997 | | | |
| **Uiso (Å$^2$)** | 0.0021(9) | | | 0.001 | | | **Uiso (Å$^2$)** | 0.0108(1) | 0.0072 |
| **ρ$_0$** | 0.080(4) | | | 0.0933 | | | **ρ$_0$** | 0.096(2) | 0.09634 |
| **R$_w$ (%)** | 7.5 | | | N/A | | | **R$_w$ (%)** | 8.9 | N/A |

**Table S3**. Crystallographic parameters of the tetragonal (model #3) and cubic (model #1) spinel models utilized in the refinements of the low-r atomic PDF (r= 1-10 Å). Reference parameters of relevant literature models are appended for comparison.

| | Bulk $Fe_{3-\delta}O_4$ | S8 | C12 | S15 Spinel | S15 Rock Salt |
|---|---|---|---|---|---|
| a (Å) | | 8.4053(1) | 8.4192(4) | 8.4419(2) | |
| b (Å) | 8.3913(2) | | | | 4.2574(6) |
| c (Å) | | 8.1950(1) | 8.2996(6) | 8.2527(2) | |
| V (ų) | 590.8643(4) | 578.9691(1) | 588.2997(1) | 588.1339(1) | 77.1673(1) |
| **Fe-Td** | | | | | |
| Uiso (Å²) at 8*b* | 0.0041(1) | 0.0030(1) | 0.0030(1) | 0.0037(1) | |
| Fe atoms/ unit cell | 9.40(1) | 6.30(4) | 7.22(1) | 7.04(1) | |
| **Fe-Oh** | | | | | |
| 1. Uiso (Å²) at 4*a* | Fe-Oh, Uiso (Å²) 0.0052(1) | 0.0029(3) | 0.0024(1) | 0.0022(2) | Fe_Uiso (Å²) 0.0029(1) |
| 2. Uiso (Å²) at 8*b* | | 0.0032(1) | 0.0027(1) | 0.0027(1) | |
| 3. Uiso (Å²) at 4*a* | | 0.0014(1) | 0.0055(3) | 0.0005(1) | |
| Fe atoms/ unit cell | 17.49(1) | 10.01(3) | 12.783(4) | 11.17(3) | |
| **O** Uiso (Å²) at 8*b* | 0.0108(1) | 0.0021(9) | 0.0028(1) | 0.0033(2) | 0.0046(6) |
| **Total Fe atoms/ unit cell** | 26.891(1) | 16.31(6) | 20.00(1) | 18.21(3) | 4 |
| $\rho_0$ | 0.096(2) | 0.080(4) | 0.085(3) | 0.082(3) | --- |
| **Refined model** | Cubic (Fd-3m) | Tetragonal (P4$_3$2$_1$2) | Tetragonal (P4$_3$2$_1$2) | Tetragonal (P4$_3$2$_1$2) | Cubic (Fm-3m) |
| **No. of phases** | 1 | 1 | 1 | 2 | |
| $R_w$ (%) | 8.9 | 7.5 | 8.4 | 7.5 | |

**Table S4**. Parameters for the nanoparticle and bulk samples, derived from fitting the low-r region of their atomic PDF (r= 1-10 Å).

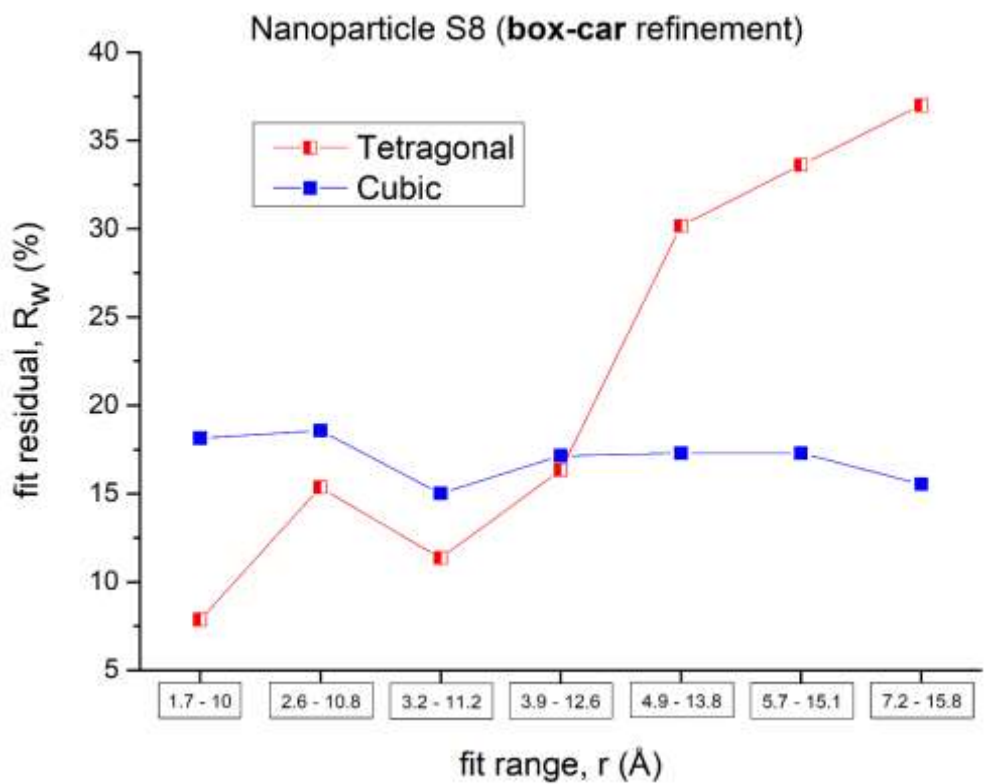

**Figure S5.** Quality of fit factors (fit residuals, $R_w$) from single-phase "box-car" refinement of the 80 K experimental G(r) data on the basis of the normal cubic spinel (model #1) and the symmetry-lowering tetragonal (model #3) models for the nanoparticle S8 sample.

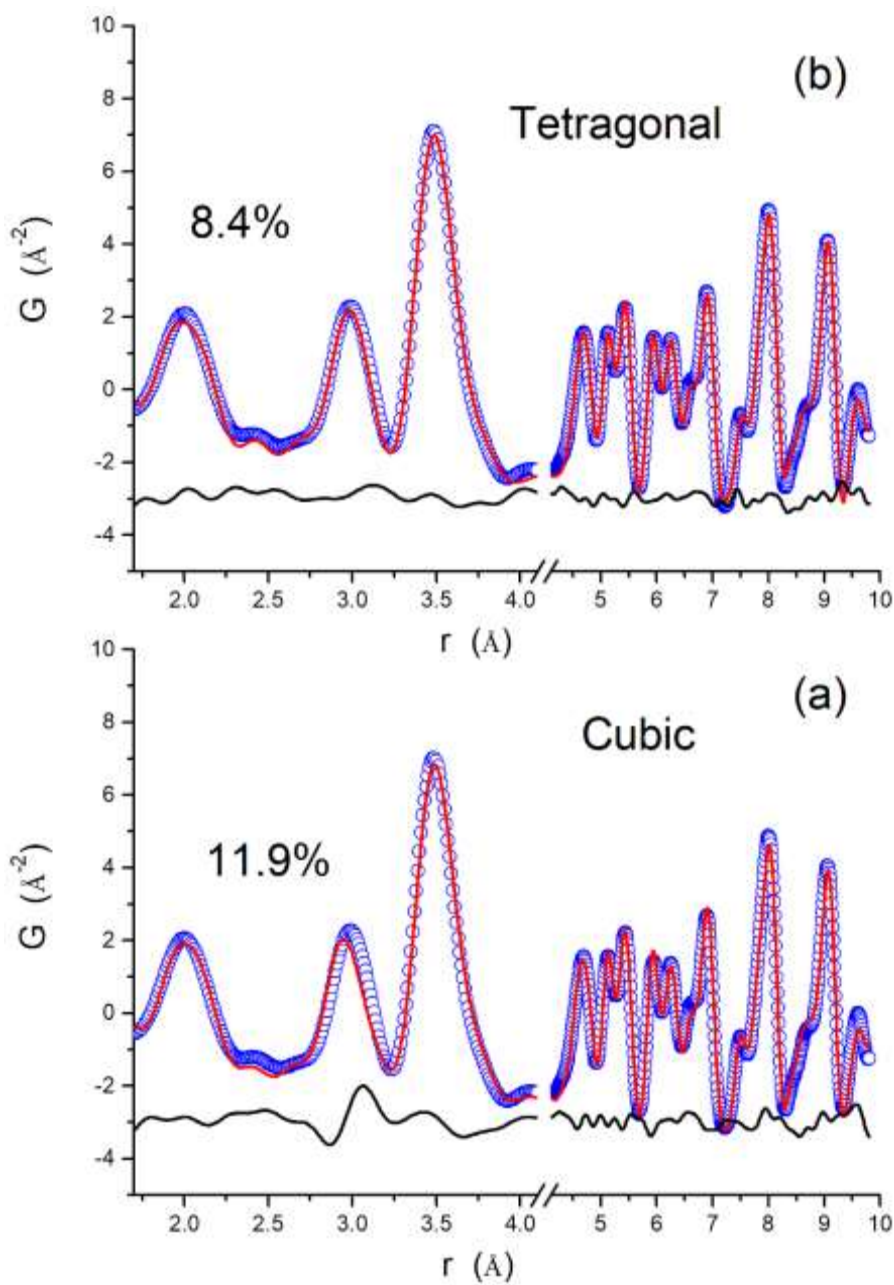

**Figure S6.** xPDF fit (T= 80 K) in the "low-r" region of the single-phase C12 nanoparticle sample assuming (a) the inverse cubic spinel atomic configuration (Fd-3m symmetry, model #1; $R_w$= 11.9%) and (b) the tetragonal (P4$_3$2$_1$2 symmetry, model #3; a$_{sp}$=b$_{sp}$=8.4192(4) Å, c$_{sp}$= 8.2996(6) Å, $R_w$= 8.4%) crystallographic configuration (see text for details). The blue circles and red solid lines correspond to the observed and simulated atomic PDFs, respectively. The black solid lines below are the difference curves between observed and calculated PDFs. The quality of fit factor, $R_w$ (%), is given for each case.

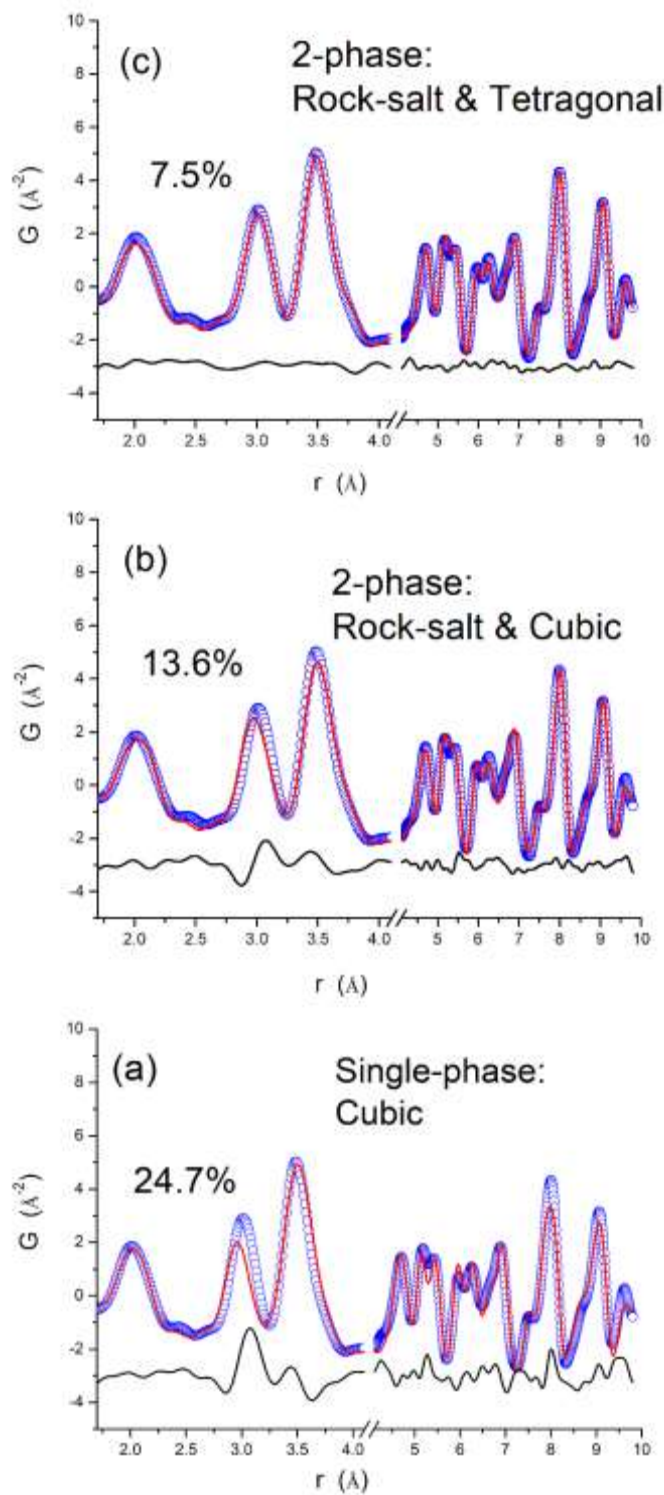

**Figure S7.** xPDF fit (T= 80 K) in the "low-r" region of the core-shell S15 nanoparticle sample assuming (a) a single phase inverse cubic spinel atomic configuration (Fd-3m symmetry, model #1), with $R_w$= 24.7%, and two-phase models of (b) the cubic rock-salt (Fm-3m symmetry) and inverse cubic spinel atomic configurations, with $R_w$= 13.6% and (c) the cubic rock-salt and tetragonal ($P4_32_12$ symmetry, model #3) crystallographic configurations, with $R_w$= 7.5% ($a_{RS}$= 4.2574(6) Å and $a_{Sp}$=$b_{Sp}$= 8.4419(2) Å, $c_{Sp}$= 8.2527(3) Å, assuming 22.5(1)%:77.5(1)% $FeO$:$Fe_3O_4$ share of volume fractions). The blue circles and red solid lines correspond to the observed and simulated atomic PDFs respectively. The black solid lines below are the difference curves between observed and calculated PDFs. The quality of fit factor, $R_w$ (%), is given for each case.

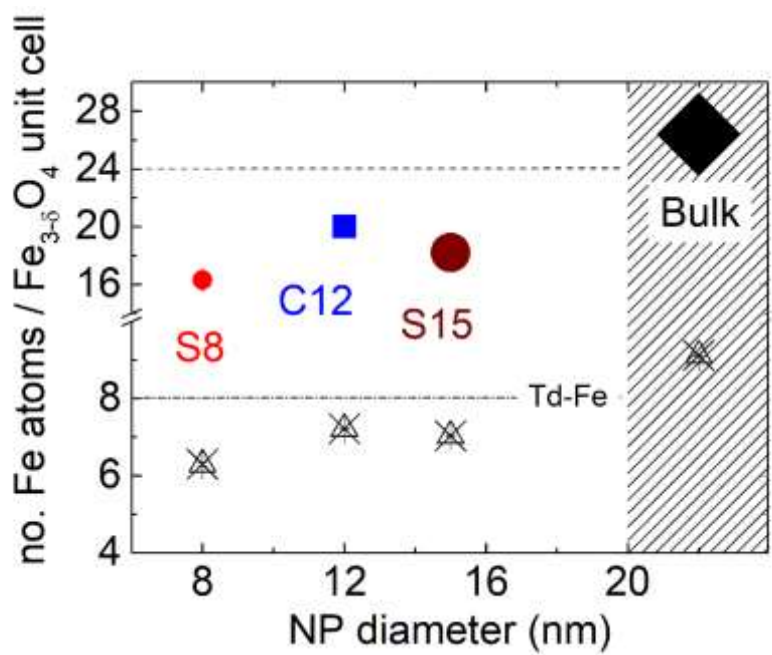

**Figure S8.** The total Fe (full symbols - standard errors are smaller than the corresponding symbol) and tetrahedral-Fe site (hatched triangles) content per unit cell for nanoparticles of different size and morphology, as extracted from the refinement of the low-r xPDF 80 K data on the basis of a single-phase tetragonal model (for S8, C12) and a two-phase model (cubic rock-salt and tetragonal configurations, for S15). These are compared against the refined Fe-contents for the bulk magnetite. The dashed line indicates the full occupancy of 24 Fe/unit cell and the dashed-dotted line the maximum occupancy of tetrahedral (Td) iron per unit cell, assuming either a cubic (or tetragonal) spinel atomic configuration.

## S8. xPDF – Fe site occupancies

xPDF data between room temperature and 80 K was modelled assuming an $Fe^{2+}$-deficient magnetite strcuture, $(Fe)_8[Fe_{1\frac{1}{3}}\langle\rangle_{2\frac{2}{3}},Fe_{12}]O_{32}$ (octahedral-Fe vacancies represented by the angular $\langle\rangle$ brackets). [3] Under this defect-based scheme randomly distributed vacancies result in an fcc lattice (model #1; normal spinel), but when their ordering is favored, either a primitive cubic ($P4_332$ symmetry, model #2) [4] structural variant could be stabilized, or a symmetry-lowering lattice distortion is triggered ($P4_32_12$ tetragonal symmetry, model #3) [2]. While model #2, similarly to model #1, also mis-calculates the 3 Å peak position, model #3 made a marked improvement in the description of the PDF peak position for both single-phase, small NPs (S8: Fig. 4d; Fig. S6b), as well as two-phase, large NPs (Fig. S7c). In the tetragonal model, in order to fit the PDF intensities accurately it was necessary to refine the Fe-site occupancies ($\eta$). Compared to the cubic spinel atomic configuration (i.e. formally, $\eta$= 24 Fe-atoms / unit cell), the refinements on the basis of model #3 indicate the following trend: $\eta$-S8 < $\eta$-S15< $\eta$-C12 [i.e. 16.4(1)< 18.2(1)< 20.0(1); Fig. S8]. That is to say, nanoparticle samples of spherical morphology display the least occupied Fe-sites. We stress that even for the cubic model #1 description of xPDF data, the Fe-sites were also given a chance to accommodate vacancies, but the symmetry lowering model #3 is a much better descriptor for the low-r xPDF data (1< $r\leq$10 Å) all the way up to room temperature.

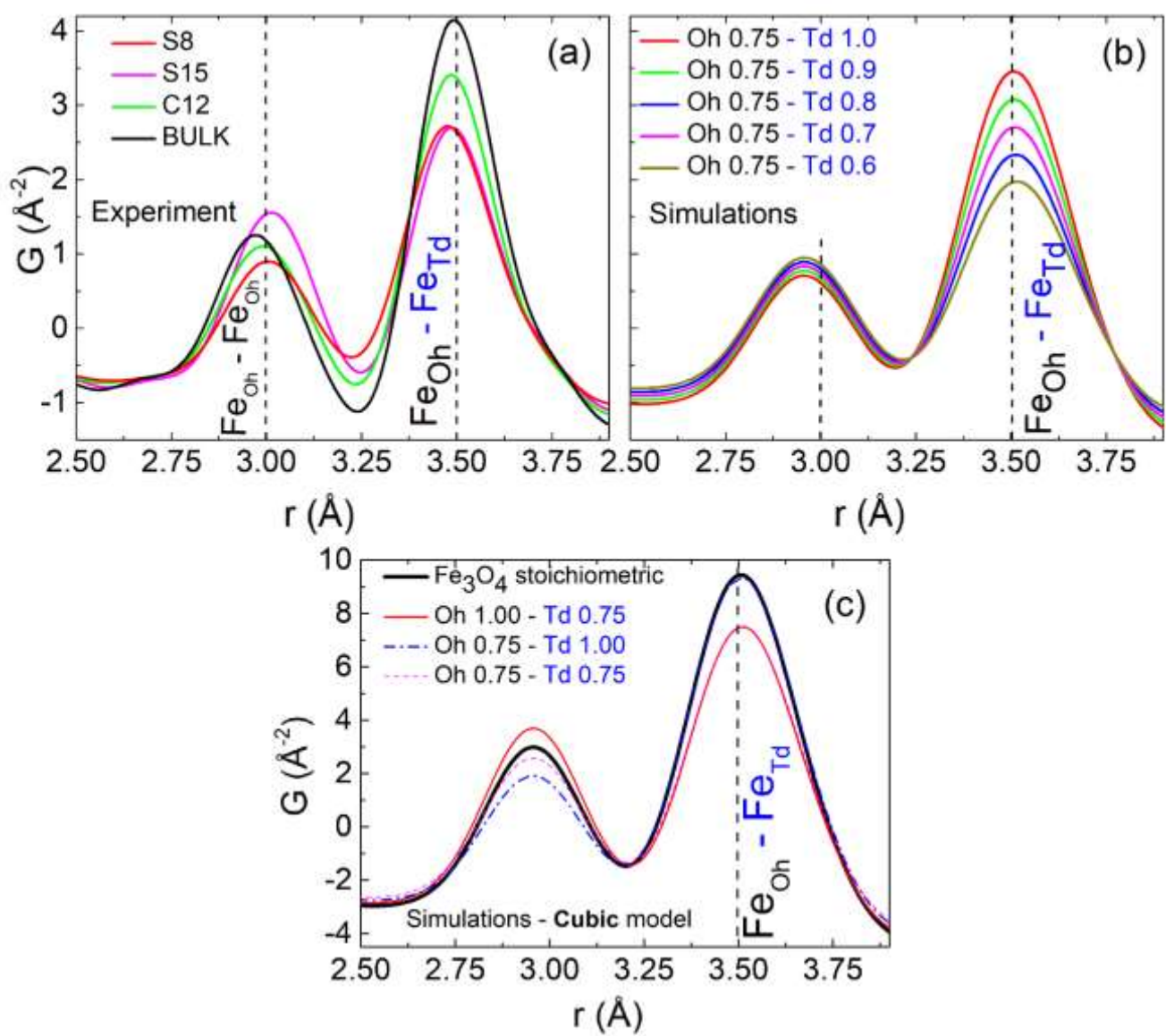

**Figure S9.** (a) Observed (T= 80 K) xPDF patterns of the low-$r$ region for all nanoparticle samples (single-phase, S8 & C12 and two-phase, S15 & C18) compared against the bulk magnetite, (b) and (c) the simulated xPDF patterns on the basis of the cubic spinel atomic configuration, where the ratio of Fe-vacancy population at the Oh and Td sites is varied. As a proof of concept in (b), the optimally chosen set of models assumes a sub-stoichiometric Oh Fe-site occupancy ($\eta$; kept constant at 75%, an average occupation derived from refinements of the G(r) raw data across the samples), while the Td Fe-site occupation level was varied in a step-wise manner (60<$\eta$<100 %). Vertical dashed lines mark the distribution of the Fe$_{Oh}$-Fe$_{Oh}$ ($r\sim$ 3.0 Å) and Fe$_{Oh}$-Fe$_{Td}$ ($r\sim$ 3.5 Å) separations from bulk to nanoscale samples.

## S9. xPDF – Where the defects are

With the purpose of evaluating whether the vacancies have a site-specific preference, xPDF patterns were simulated while the ratio of Fe-vacancy population at the Oh and Td sites was varied (selected occupancies were quantified through refinements). The trend is similar assuming either the symmetry-lowering model #3 (Fig. 5) or the cubic spinel model #1 (Fig. S9).

(a) With vacancies residing at the Oh sites alone, simulations indicate that the G(r) peak-intensity at 3.5 Å is not suppressed, while that at 3.0 Å is somewhat reduced with respect to bulk (Fig. S9c).

(b) However, when Fe-vacancies are accommodated only at the Td sites, the G(r) at 3.5 Å is diminished significantly with respect to bulk, while a small increase at the at 3.0 Å peak intensity is observed (Fig. S9c).

(c) In the event that both Oh and Td sites are deficient to a similar degree, the impact of sub-stoichiometry (due to the counter-balance of the scattering power at the two sites) on the G(r) is rather hard to conclude (Fig. S9c).

Intrinsic lattice defects, commonly met in the form of octahedral Fe-vacancies mediate the structure and properties of spinel iron oxides, [5] but the progressive intensity diminution at $r \sim 3.5$ Å, when raw and simulated patterns are compared (Fig. 5, Fig. S8) corroborates to a significant volume of vacancies also at the Td Fe-sites (~10-20%; Fig. S8). Cationic vacancies, at both Oh and Td Fe lattice positions in related $\gamma$-Fe$_2$O$_3$ colloidal NCs, assessed before by xPDF analysis, inferred a limited length of structural coherence in NCs [6] as compared to bulk samples where Fe-vacancies reside primarily at Oh sites. The significant content of vacancy distribution, as a consequence of the oxidation mechanisms of wüstite, reflects also in the cell dimensions and their temperature evolution (Fig. S10, §S10). The latter point to strains/stresses, observed also in the single-particle local structure (probed by STEM-HAADF imaging, electron tomography, and holography), [7], [8] both at the core@shell interface and the spinel shell itself of Fe$_x$O-Fe$_3$O$_4$ core@shell nanocubes.

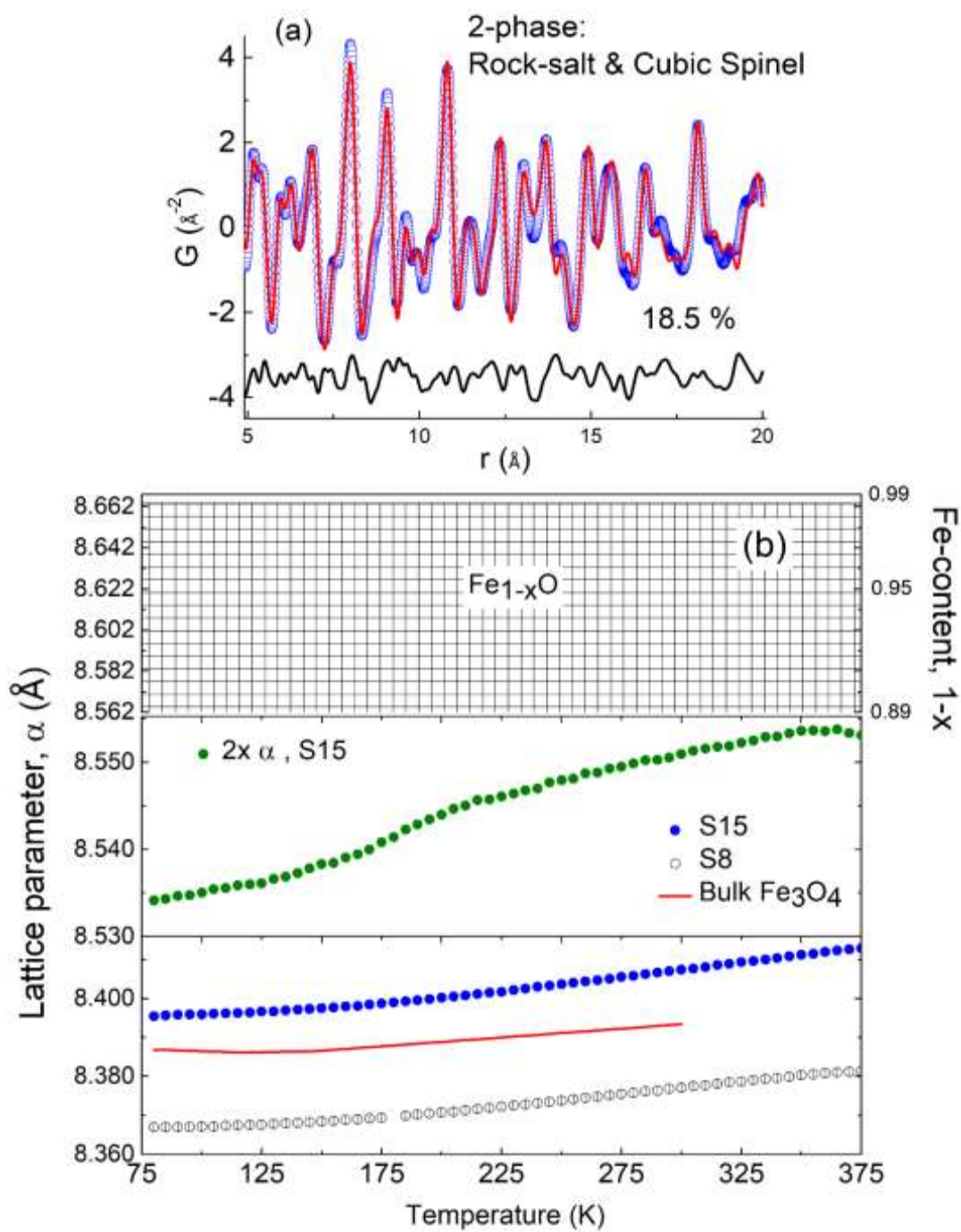

**Figure S10.** (a) xPDF fit (T= 80 K) in the higher-r region of the core-shell S15 nanoparticle sample assuming a two-phase model of cubic rock-salt (Fm-3m symmetry) and cubic spinel atomic configurations, with a quality of fit factor, $R_w$= 18.5%. (b) The temperature evolution of the lattice parameters for the spherical morphology nanoparticles (NPs) of 8 and 15 nm against those of bulk magnetite; standard errors are smaller than the corresponding data symbol. The cell dimensions were derived from the xPDF refinements in the r-range away from one unit cell (r= 5 – 20 Å). The fittings were carried on the basis of cubic spinel crystallographic model for the bulk $Fe_3O_4$, as well as for the small NPs (S8), but a two-phase cubic rock-salt and spinel atomic configurations (FeO - $Fe_3O_4$) was used for the larger core-shell S15 NPs. The cell dimensions for the bulk $Fe_3O_4$ (Fd-3m symmetry; model #1) is represented by the red line. The hatched region indicates the average lattice parameters for, defect rock-salt $Fe_xO$ bulk samples met in literature; the rock-salt lattice dimensions have been doubled for ease of comparison against the cubic spinel magnetite.

**S10. Cell-size *T*-evolution – Average structure**

Clear deviations from the ideal cubic spinel atomic configuration are recognizable already at room temperature during the analysis of the local structure (1-10 Å), however, refinements of the xPDF at r> 10 Å point that there is no much structural coherence of the underlying tetragonal distortion beyond one unit cell (R-factors of the order of 35-40 %). Then one may ask whether the symmetry lowering distortion could average out to cubic over some longer length-scale. For this reason the refinement of the xPDF data at relatively longer radial distances (5 < r < 20 Å) was carried out by assuming either the cubic spinel configuration for the smaller NPs (S8, C12) or the two-phase rock-salt and cubic spinel (FeO - $Fe_3O_4$) models for the larger core-shell NPs (S15). Analysis on the basis of such cubic structural models points that there is observable mismatch in the peak positions (as such, R-factors cannot become a lot better than 19-20 %; Figure S9a), indicating that spatially averaged distortions are poorly described by the average structure cubic model. Per present analysis and assessments of fits over broader r-ranges it appears that these nanoscale samples are highly non-uniform in terms of defects/vacancy distributions, generating strains, and their structure is not quite cubic at high-r, and is neither so locally too.

In this framework, we attempted to evaluate the impact of the local effects on the structural coherence beyond one unit cell (r= 5-20 Å) by comparing the cubic lattice constants of each iron oxide phase in the spherical NPs (S8 and S15) against those of the bulk magnetite (Fig. S10b). The fully oxidized S8 sample adopts a contracted spinel-like lattice, while the equivalent type of phase in the shell of the S15 NPs is somewhat expanded with respect to the bulk. The latter is accompanied by the contraction of the $Fe_xO$ rock-salt type of core in the S15. The apparent expansion of the shell against the contraction of the core in the S15, is a likely consequence of the system's effort to optimize its elastic energy gain when the heterostructured (cf. ~3% mismatch between the fcc cells of spinel, $a_{0-Fe3O4}$ = 8.39 Å, and rock-salt, $a_{0-FeO}$ = 4.29 Å, type of structures) particle is formed during self-passivation.

In effect, the evaluation of the *T*-dependence of the cell dimensions provides an insight about the effect of Fe-vacancies (§1.2.2) and the subsequent relaxation of the average crystal structure in order to accommodate them in the cell. Overall, refinements of the xPDF data (r= 5-20 Å) tell us that in the S15 NPs the $Fe_xO$ core carries compressive strain instead of the tensile one at the shell, in line with earlier GPA analysis [9] of $Fe_xO$-$Fe_3O_4$ core-shell nanocubes. Moreover, as the lattice parameter of the defect rock salt structure of $Fe_xO$ varies linearly with x over the entire compositional range, [10], [11] the average stoichiometry of the wüstite core in the S15 is estimated as $Fe_{0.88(1)}O$ (a= 4.2756(2) Å, refined from xPDF, at 300 K). This Fe-deficient content is likely to alleviate the lattice-mismatch (cf. relieved to ~1.8% for the confined $Fe_{0.88(1)}O$ under the present growth method) between rock-salt and spinel phases, thus aiding the core-shell topological phase formation.

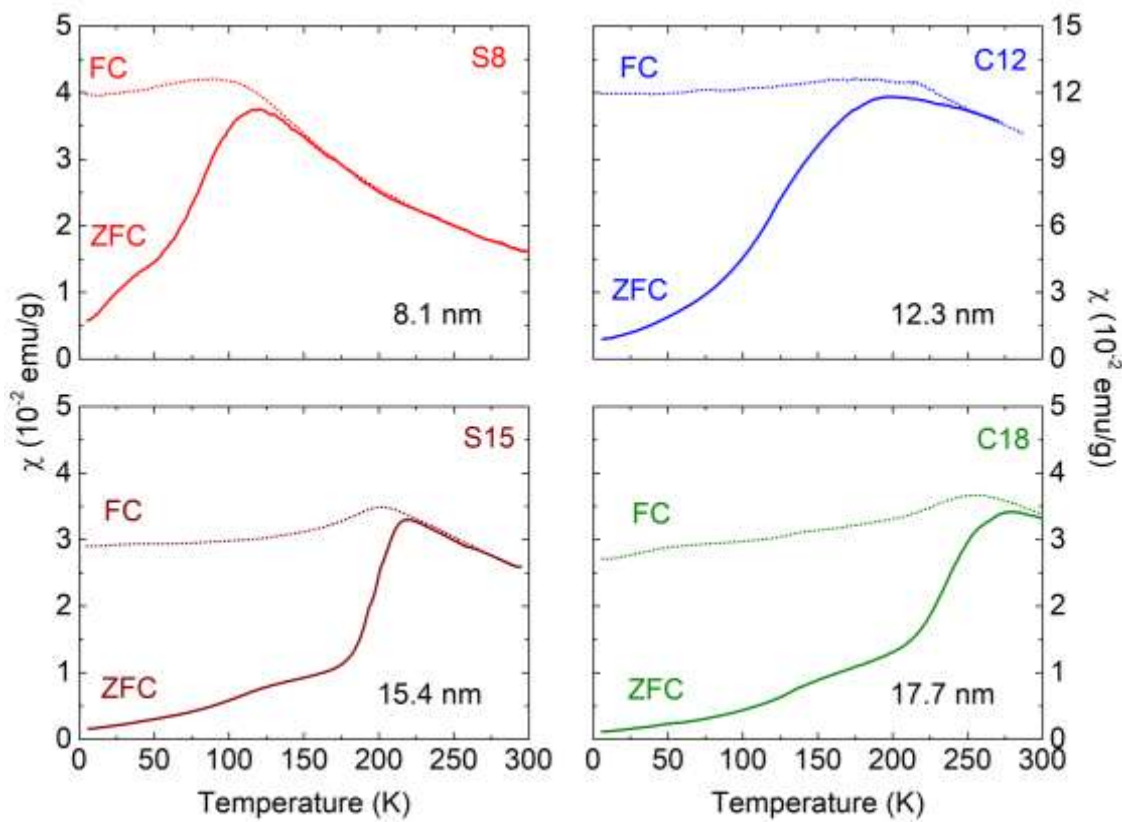

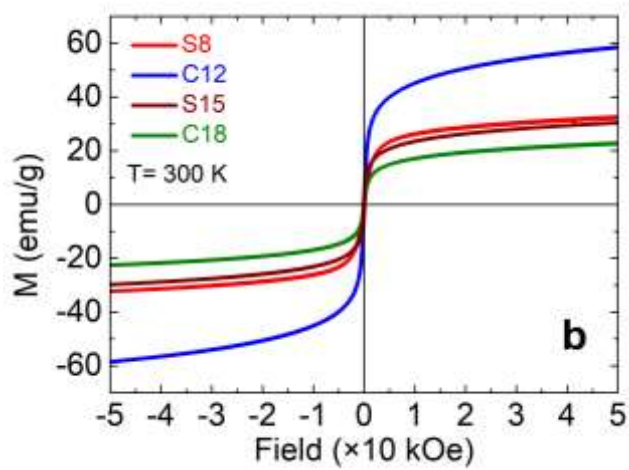

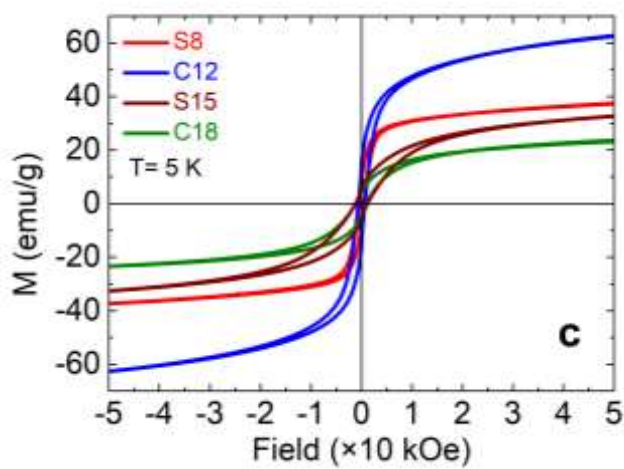

**Figure S11.** (a) The temperature evolution of the zero-field cooled (ZFC; solid lines) and field-cooled (FC; dotted lines) susceptibility curves for the single-phase S8, C12 and core@shell S15, C18 nanoparticles under a magnetic field of 50 Oe. The hysteresis loops at 300 K (b) and 5 K (c), for the spherical (S8, S15) and cubic (C12, C18) nanoparticles.

## S11. Effective magnetic anisotropy

Assuming $T_B$ *is* approximated by the maximum in ZFC curves, then from the simplified Néel-Brown equation $K_{eff} V = 25k_B T_B$ [12] for a single magnetic domain, $K_{eff}$ is expected to increase linearly with $T_B$ at constant volume. Details in spin reversal processes can be further masked by particle-size distributions, and importantly by structural defects that can give rise to non-uniform spin rotations. Thus, values of the $K_{eff}$ estimated on the basis of the Néel-Brown model alone are likely misjudged here. In real systems, anisotropy dictated by size, surface and magnetic exchange coupling at interfaces, correlates with structural variations amongst the NPs. Additional contributions should be explicitly taken into account, namely $K_{eff} = K + (6\,K_s/d)$ [13], where $K_s$ is the surface anisotropy (e.g. reduced coordination due to broken surface bonds tend to boost the anisotropy of fine particles) [14], [15] and $d$ is the particle diameter, while the volume anisotropy, $K$, may consist of extra terms too, such as the magnetocrystalline, magnetostriction and shape anisotropies. These local changes of the magnetic anisotropy can destabilize the collinear spin arrangement of the Stoner-Wohlfarth model and give rise to a complex internal spin structure. [16,17]

We note that the experimentally determined low temperature coercive fields ($H_C$: strongly relates to the magnetic anisotropy and domain structure) of the smaller NPs (Table S3) are larger than that expected (~75 Oe) for the coherent magnetization reversal of randomly oriented particles, [18] thus suggesting that the process is controlled by a different mechanism, whose strength is larger than that of the magnetocrystalline anisotropy alone. [19] The observation of the exchange anisotropy field in the field-cooled M-H curves comes into further support of the previous argument. Moreover, the Monte Carlo simulations corroborate that at subcritical particle sizes (≤12 nm) defect-induced disorder and spin-frustration (i.e. competing FiM and AFM interactions) at surfaces and interfaces introduce greater complexity in the potential energy landscape than that (of the simple double-well potential) for coherent spin reversals, predicted by the Neel-Brown theory. On the basis of these observations, we are of the opinion that $K_{eff}$ determined via $T_B$ from the present macroscopic magnetic properties, is an indication test. The progressive increase of the measured $T_B$s should be considered as a manifestation of the competition between raising exchange anisotropy at interfaces and the temperature dependent anisotropy of the surrounding spinel-like phases, [20] corroborating that processes beyond the coherent reversal of $M$ are involved in these spontaneously self-passivated wüstite NPs.

| Sample Label | Size (nm) | $T_B$ (K) | $M_S$ (emu/g), 300 K | $H_c$ (Oe), 300 K | $M_S$ (emu/g), 5 K | $H_C$ (Oe), 5 K | $M_S$ (emu/g), FC curve, 5 K | $H_C$ (Oe), FC curve, 5 K | $H_{EB}$ (Oe) |
|---|---|---|---|---|---|---|---|---|---|
| S8 | 8.1 | 119 | 32.3 | 101 | 37.3 | 523 | 37.8 | 545 | 251 |
| C12 | 12.3 | 198 | 58.4 | 86 | 62.5 | 698 | 62.8 | 589 | 200 |
| S15 | 15.4 | 218 | 30.0 | 177 | 32.5 | 1246 | 33.0 | 1589 | 1050 |
| C18 | 17.7 | 277 | 22.7 | 145 | 23.5 | 1184 | 23.5 | 1368 | 725 |

**Table S5.** Geometric characteristics of self-passivated Fe$_x$O-Fe$_3$O$_4$ nanocrystals, their blocking temperatures, $T_B$, as well as the saturation magnetization, $M_S$, and the coercive field, $H_c$ at T= 5 and 300 K. The field-cooled (FC) hysteresis loop (*M-H*; $H_{cool}$= 50 kOe) characteristics, including exchange bias, $H_{EB}$, at 5 K are summarized.

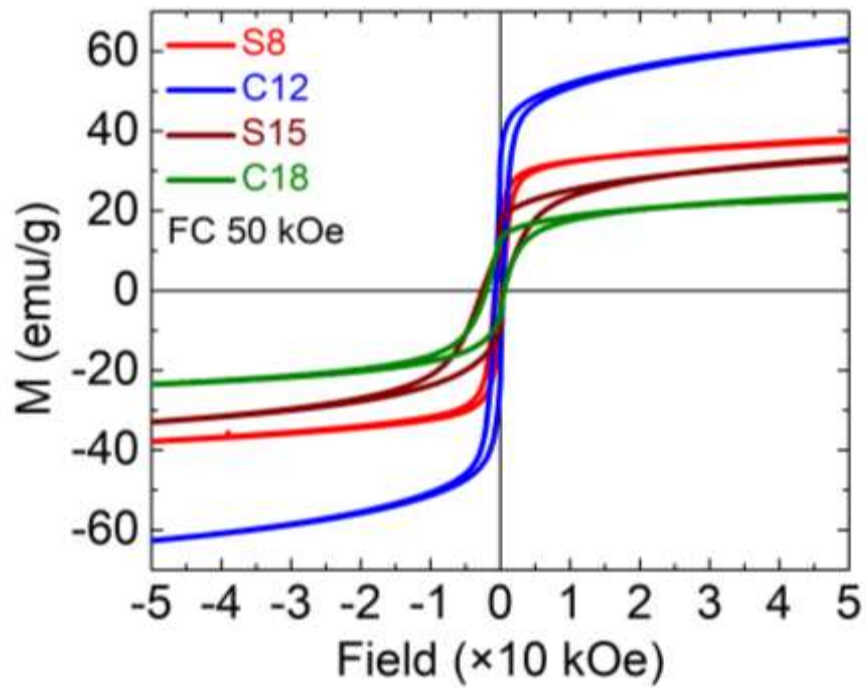

**Figure S12.** The hysteresis loops at 5 K under field-cooling of 50 kOe, for spherical (S8, S15) and cubic (C12, C18) nanoparticles.

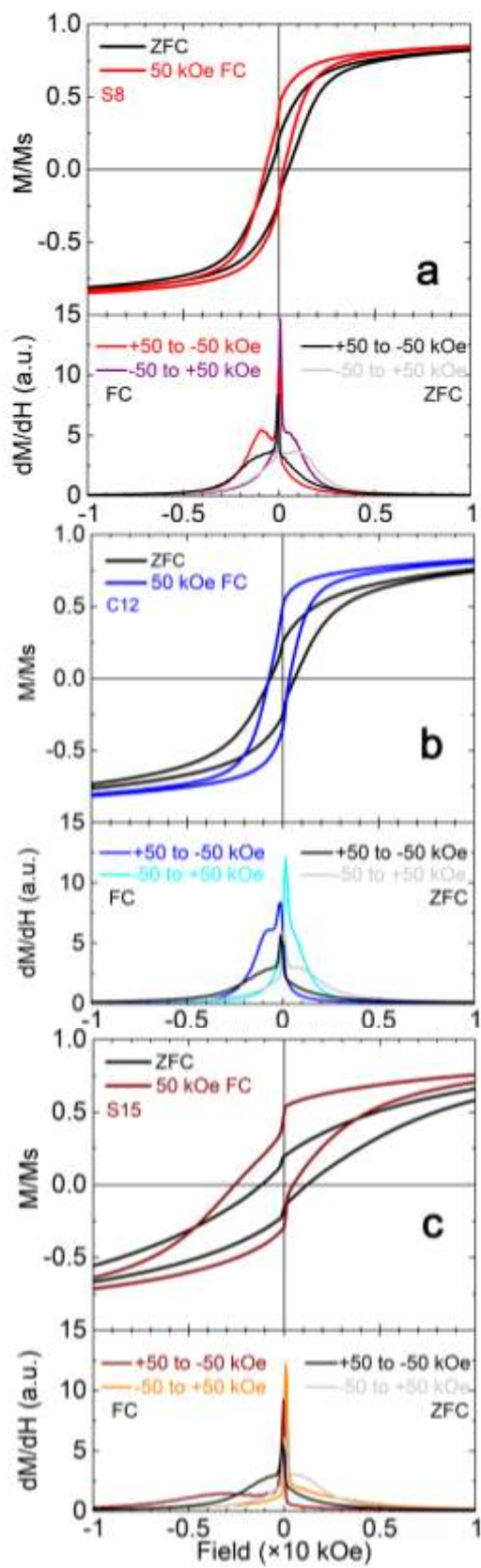

**Figure S13.** The low-field, normalized hysteresis loop regions at 5 K, for the single-phase S8 (a), C12 (b) and the core@shell S15 (c) nanoparticles. It compares the M(H)/M$_s$ data measured after a 50 kOe field-cooling protocol (FC; colored curve) and the corresponding M(H) data taken under zero-field cooled (ZFC; black curve) conditions. Each panel beneath the normalized magnetization M/M$_s$ data presents the corresponding derivative curves from +ve to –ve field sweeps and the vice-versa.

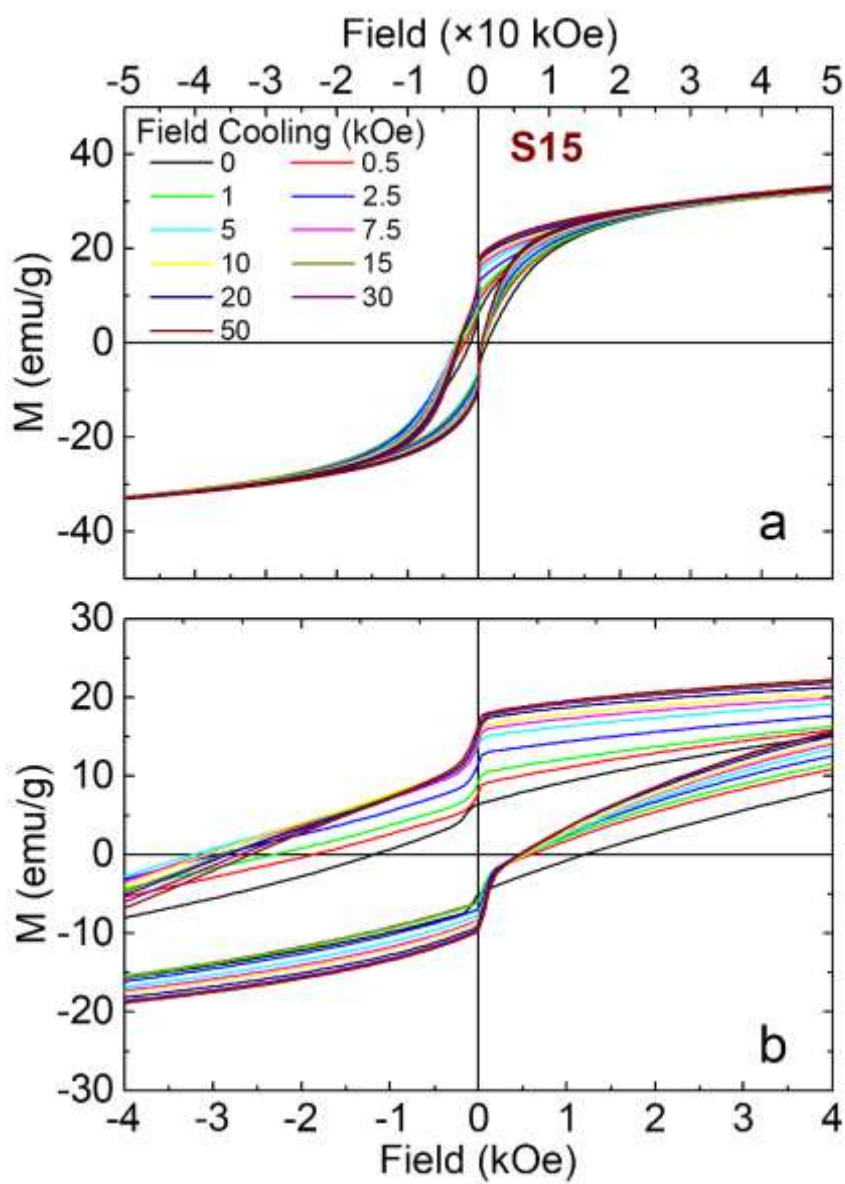

**Figure S14.** Full (a) and partial (b) magnetization hysteresis loops for the spherical core-shell nanoparticles (S15) obtained after field-cooling at progressively elevated applied fields ($0 \leq H_{cool} \leq 50$ kOe).

## S14. Exchange anisotropy

As granular exchange-coupled nanoscale systems display an interesting dependence on the cooling-field strength ($H_{cool}$), [21] the M(H) behaviour of the S15 NPs was further investigated (Fig. S14). The quick rise of $H_{EB}$ (Fig. 7; left y-axis) and its significantly larger magnitude against the single-phase S8 and C12 is clearly resolved. At $H_{cool} >5$ kOe, the Zeeman coupling partially wins over the biasing effect that turns the spins of the shell less strongly coupled to the AFM core, rendering $H_{EB}$ somewhat reduced and saturated (Fig. 7). [22], [23] In such topologies, the buried FiM shell interface promotes the competition of the exchange interactions by establishing an increased fraction of uncompensated AFM core spins [24] that boost the $H_{EB}$ (and $H_c$) in the S15 NPs.

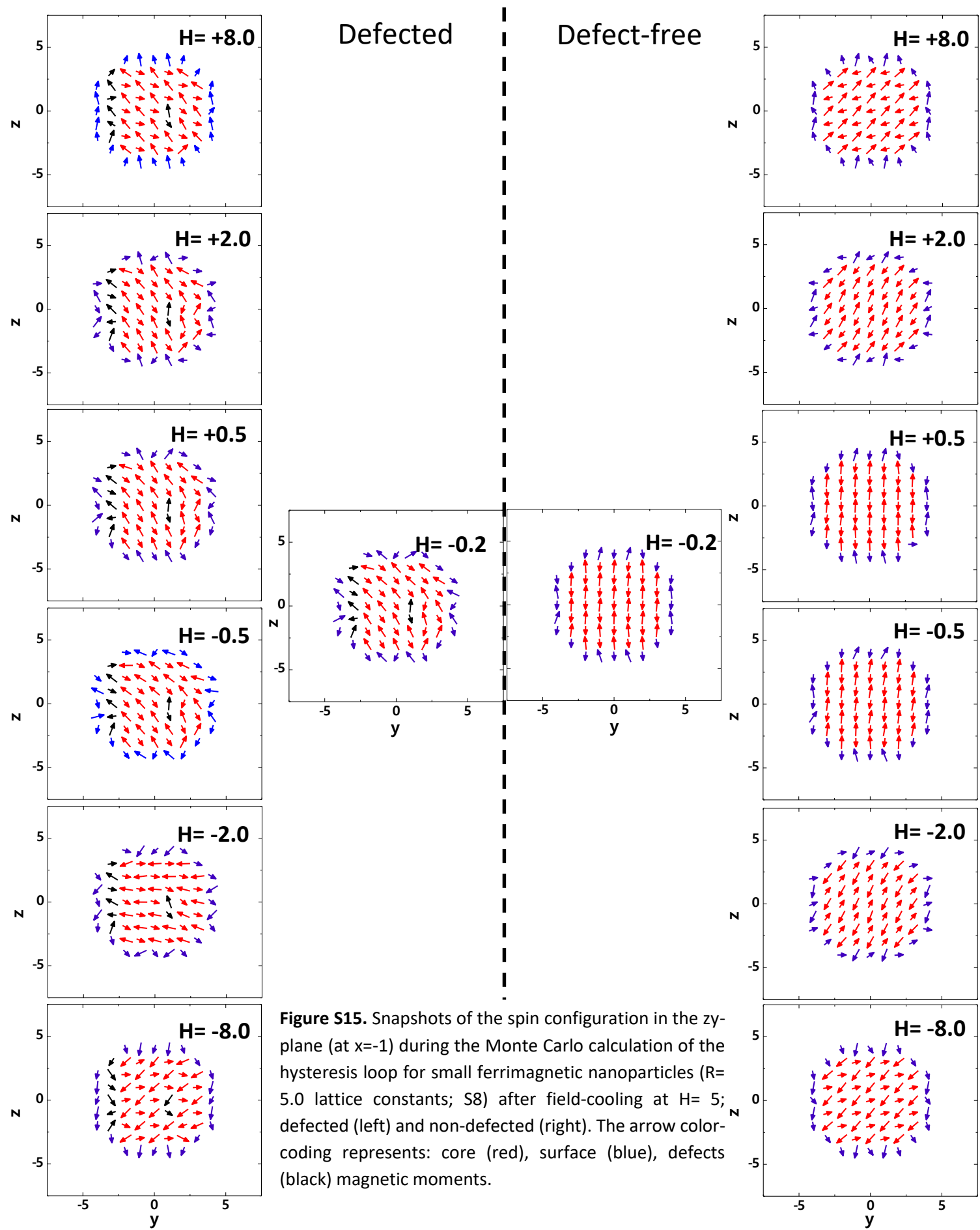

**Figure S15.** Snapshots of the spin configuration in the zy-plane (at x=-1) during the Monte Carlo calculation of the hysteresis loop for small ferrimagnetic nanoparticles (R= 5.0 lattice constants; S8) after field-cooling at H= 5; defected (left) and non-defected (right). The arrow color-coding represents: core (red), surface (blue), defects (black) magnetic moments.

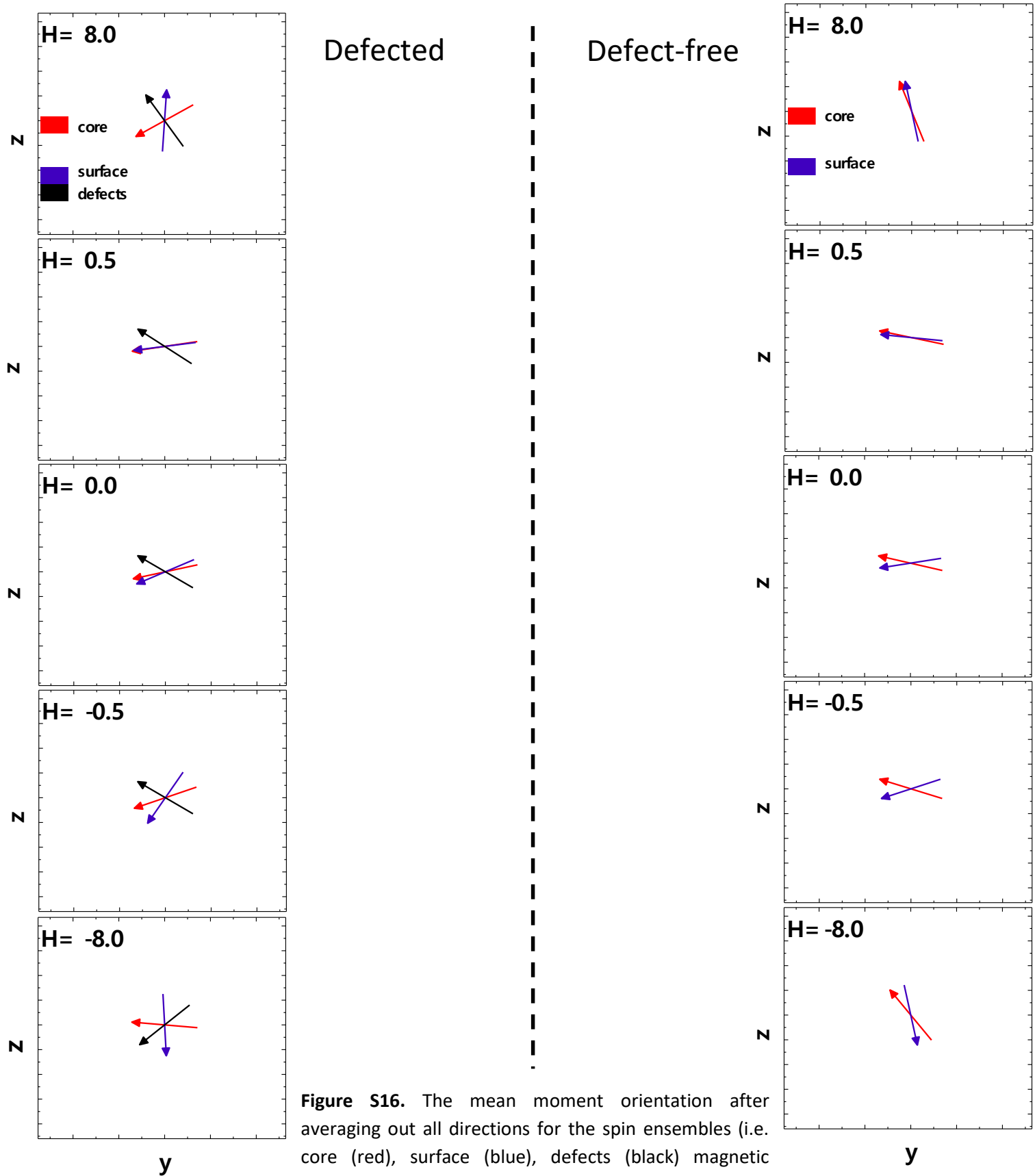

**Figure S16.** The mean moment orientation after averaging out all directions for the spin ensembles (i.e. core (red), surface (blue), defects (black) magnetic moments) encountered in the zy-plane (x= -1) during the Monte Carlo calculation of the hysteresis loop for small ferrimagnetic nanoparticles (R= 5.0 lattice constants (S8)).

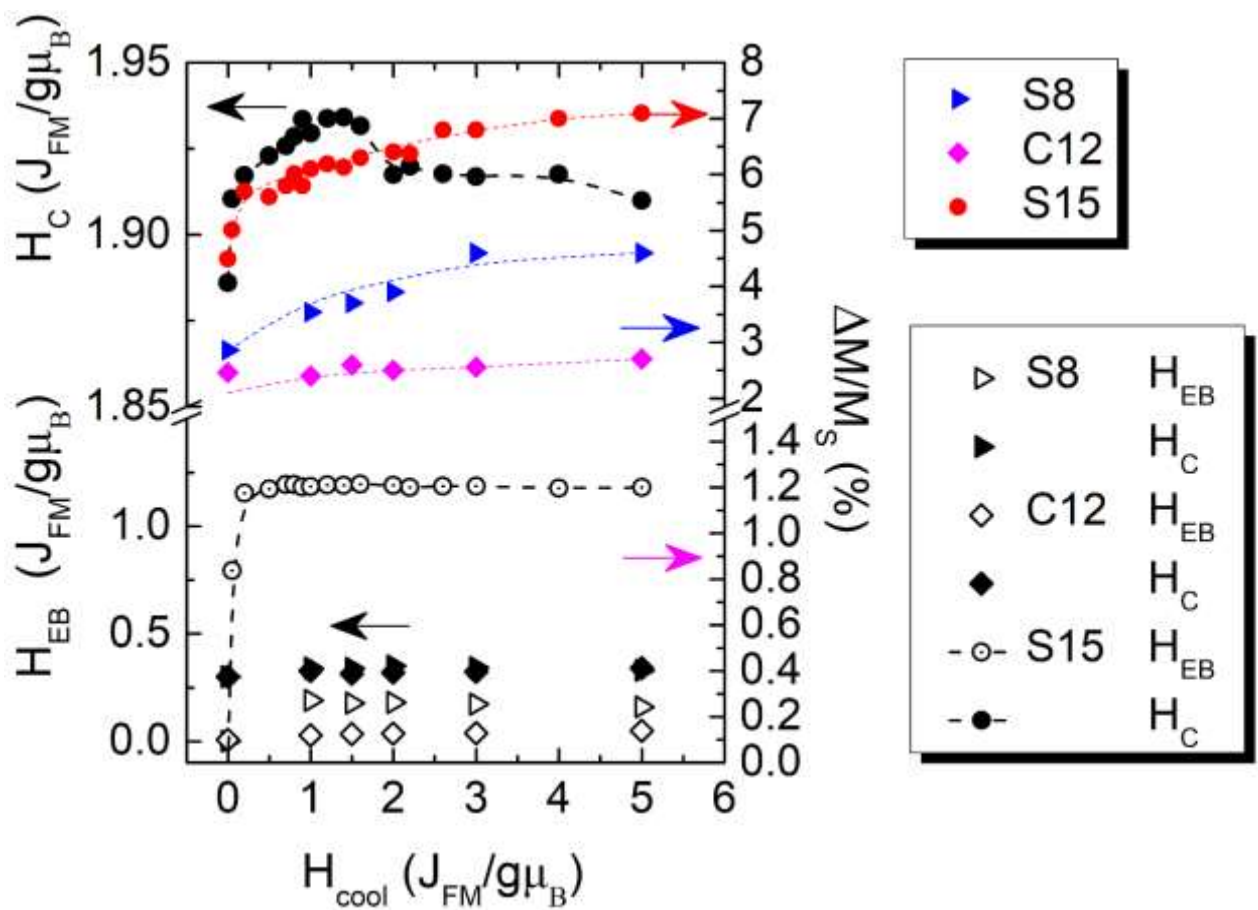

**Figure S17.** Monte Carlo theoretical calculation of the exchange-bias ($H_{EB}$), coercive field ($H_C$) and low-field jump ($\Delta M/M_S$) obtained at varying cooling-field strengths ($H_{cool}$), for spherical core-shell (S15) and single-phase (S8, C12) nanoparticles, with geometric characteristics similar to those studied in the experiments.

### S17. Monte Carlo calculation of hysteresis loop parameters for large core-shell NPs

When similar type of defects is introduced in the larger S15 NPs, as they are distributed both in the core and the shell sub-domains, the MC simulations indicate that $\Delta M/M_S$ initially grows (Fig. 8; left y-axis), pointing to an extended two-phase character, but then no further changes occur. A small enlargement of $H_{EB}$ was also observed (Fig. 8; right y-axis), originating from processes similar to those discussed for the core in the smaller C12 and S8 NPs, but strong interface anisotropy due to AFM/FiM lattice mismatch has a leading role now. Within this model, as magnetization states at the different kinds of interfaces are adjusted by $H_{cool}$, MC simulations (Fig. S17) reproduce fairly well the evolution of the hysteresis loop parameters ($H_{EB}$, $H_C$, $\Delta M/M_S$) for the S15 NPs.

**S18. Monte Carlo calculation of the Specific Absorption Rate (SAR): Defected *vs.* Defect-free NPs**

Heterogeneity in apparently chemically uniform small NPs ($\leq$10 nm), as manifested, for example, by the bare surface to inner atoms volume ratio, is an important property-tweaking parameter. Here, we implement the Monte Carlo (MC) simulations technique to calculate the SAR for a defected nanoparticle system derived from the self-passivation of $Fe_xO$-$Fe_{3-\delta}O_4$ nanocrystals. The particle assumes a magnetite-like phase with a size of D= 8.1 nm (S8), and a surface layer thickness of $Fe_3O_4 = 0.8397 \times 10^{-9}$ m.

The MC calculation was undertaken on the basis of the approximation set by a modified linear Néel-Brown relaxation model (susceptibility losses) [25] for core(soft)-surface(hard) ferrimagnetic NPs. [26] The SAR due to susceptibility losses is expressed as: $SAR(f) = \frac{\mu_0 \pi f \chi'' H_0^2}{\rho}$

$\rho$: average density of each ferrite nanoparticle: $\rho_{Fe3O4} = 5.24 \times 10^3$ kg/m$^3$,

$H_0$ : AC field amplitude, and $f$ : field frequency,

$\chi''$: imaginary part of the complex susceptibility that involves the effective relaxation times for the two absorption mechanisms, namely, Brown ($\tau_B$) and Néel ($\tau_N$).

Temperature is set to T= 313 K.

$\eta$ : the medium viscosity with a value of 0.65 x $10^{-3}$ Pa·s (approximately the value for water at 40°C)
$\phi$ : volumetric ratio of the NPs set to 0.001

A surfactant layer that covers the nanoparticles (NPs) is taken to have a thickness of 4 nm that is a parameter introduced in the calculation for the Brownian relaxation time.

Taking into account that the effective nanoparticle anisotropy constant is given as $(K V)_{eff} = K_{core} V_{core} + K_{surf} V_{surf}$ and the experimental $K_{eff}$=14.8 $10^4$ J/m$^3$, we take $K_{core}$= 7.41 $\times 10^4$ J/m$^3$ and $K_{srf}$ = 3 $\times$ $K_{core}$ based on the simulated parameters.

The experimental saturation magnetization for S8 was $M_S$= 1.69 $\times 10^5$ A/m. We then calculated the effective $(MV)_{eff}$= $M_{core} V_{core}$+ $M_{surf} V_{surf}$ , where the saturation magnetizations of the core and the surface are set equal to the product of the MC calculated values of the normalized corresponding magnetizations, Mi(MC) $\times M_S$.

For comparison purposes, we considered magnetite NPs of similar diameter, D= 8 nm that are assumed non-defected in their core. [27]

Here, the experimental anisotropy was estimated as $K_{eff}$=2.73 $\times 10^4$ J/m$^3$ and $M_S$=2.62 $\times 10^5$ A/m, so we take the $K_{core}$= 1.36 $\times 10^4$ J/m$^3$ and $K_{srf}$ = 3 $\times$ $K_{core}$

While exploring the effect of AC field strength, the SAR according to the Linear Response Theory for the Néel-Brown relaxation model, [25] was calculated at a field amplitude, $H_0$, with a frequency of $f$= 500 kHz. The difference in the magnitude of SAR (due to susceptibility losses) between defected and defect-free NPs is depicted in the plot shown in Fig. S18. The SAR ($H_0$) curves follow a trend similar to that in the experimental findings. [28]

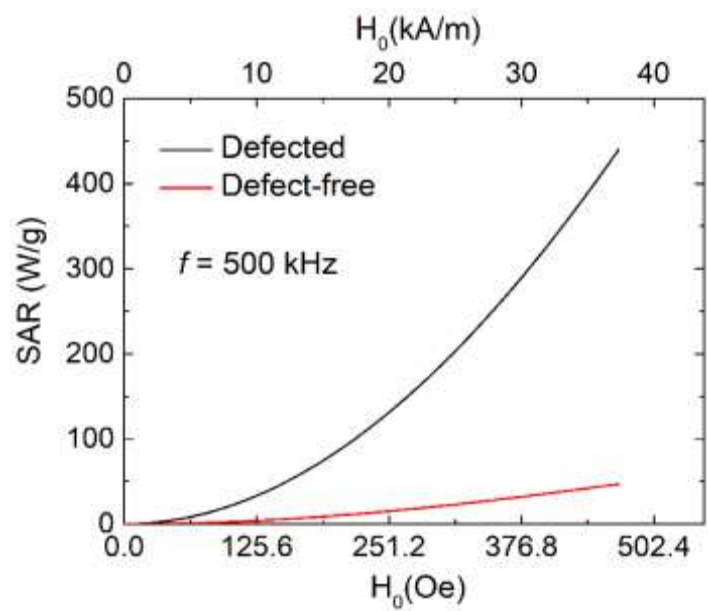

**Figure S18.** MC calculation of SAR values as a function of the AC field amplitude, $H_0$ at a field frequency of $f$ = 500 kHz.

In summary, the outcome of the MC simulations is that in the case of small defected nanoparticles, where the effective anisotropy increases 5 times compared to the defect-free ones, the SAR values are raised almost ten-fold at a frequency of 500 kHz and an applied field amplitude $H_0$= 37.3 kA/m .

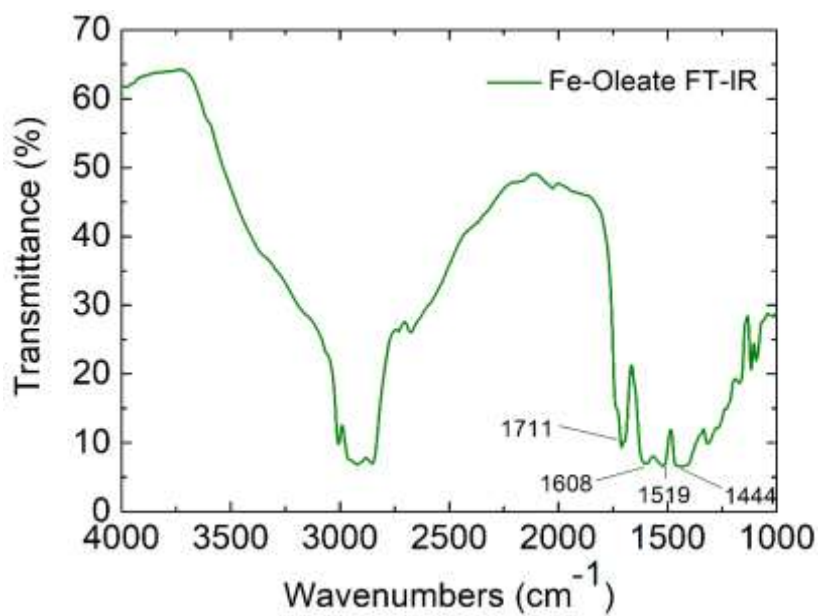

**Figure S19.** FTIR spectrum of Fe-Oleate.

**S20. Methods - Monte Carlo Simulations**

The simulations considered three nanoparticle model systems, with somewhat different morphological features as indicated by the experimental observations. Initially, large S15 spherical nanoparticles (NPs) of an average radius, R= 9.1 and a shell thickness of 4 lattice spacings (expressed in terms of the simple cubic cell of magnetite, a= 8.39 Å), were assumed to entail an AFM core and a FiM shell structure, resembling that in the S15 specimen. The spins in the NPs interact with nearest neighbor Heisenberg exchange interaction, and at each crystal site they experience a uniaxial anisotropy based on previous theoretical [12] and experimental [29], [9], [30] knowledge. Under an external magnetic field, the energy of the system is calculated as: [22], [24]

$$E= -J_{core} \sum_{i,j \in core} \vec{S_i} \cdot \vec{S_j} - J_{shell} \sum_{i,j \in shell} \vec{S_i} \cdot \vec{S_j} - J_{IF} \sum_{i \in core, j \in shell} \vec{S_i} \cdot \vec{S_j}$$

$$-K_{i \in core} \sum_{i \in core} (\vec{S_i} \cdot \hat{e}_i)^2 - K_{i \in shell} \sum_{i \in shell} (\vec{S_i} \cdot \hat{e}_i)^2 - \vec{H} \sum_i \vec{S_i} \qquad (1)$$

Here $\vec{S_i}$ is the atomic spin at site $i$ and $\hat{e}_i$ is the unit vector in the direction of the easy-axis at site $i$. We consider the magnitude of the atomic spins in the two AFM sublattices equal to 1 and in the two FiM sublattices of the shell to be equal to 1 and 1.5, respectively. The first term in eq. 1 gives the exchange interaction between the spins in the AFM core, while the second term gives the exchange interaction between the spins in the FiM shell. To take into account the difference of the magnetic transition temperatures [$T_N$(core) <$T_C$(shell)], we consider the exchange coupling constant of the core as $J_{core}$ = -0.1 $J_{FM}$ and that of the shell as $J_{shell}$ = -1.5 $J_{FM}$, where $J_{FM}$ is considered to be the exchange coupling constant of a pure ferromagnet (FM), $J_{FM}$= 1, a reference value. The third term gives the exchange interaction at the interface between the core and the shell. The interface includes the last layer of the AFM core and the first layer of the FiM shell. The exchange coupling constant of the interface $J_{IF}$ is taken between the value of the core and the shell $J_{IF}$= -0.3 $J_{FM}$. The fourth term gives the anisotropy energy of the AFM core, $K_C$. If the site $i$ lies in the outer layer of the AFM core $K_{i-core}$ = $K_{IF}$ = 5 (due to strong lattice mismatch) and $K_{i-core}$ = $K_C$ = 0.5 $J_{FM}$ elsewhere. The fifth term gives the anisotropy energy of the FiM shell, which is taken as $K_{shell}$ = 0.1 $J_{FM}$. If the site $i$ lies in the outer layer (i.e., the surface) of the shell then the anisotropy is taken as $K_{i-shell}$= $K_S$= 1.0 $J_{FM}$, which is assumed to be random (rather than uniaxial). The last term in eq. 1 is the Zeeman energy.

In the case of smaller S8 single-phase spherical nanoparticles, with R= 5 lattice constants, the interface terms were waived out, while the anisotropy parameters were taken as $K_C$ = 0.1 $J_{FM}$, with $J_{core}$ = -1.5 $J_{FM}$ for the core contributions (similarly to the case of the FiM shell), while for the surface anisotropy and associated exchange coupling, $K_S$= 0.3 $J_{FM}$, with $J_{shell}$= -0.1 $J_{FM}$, respectively. When single-phase cubic nanoparticles with R= 7 (C12) lattice constants were investigated, the relevant core anisotropy parameters were the same as for the single-phase spherical NPs, but $K_S$= 0.15 $J_{FM}$ was chosen to be weaker than that of the spherical counterparts due to the less significant surface effects. In all the aforementioned cases, the impact of structural defects was also taken into account. For this purpose the modeling approximated them as randomly distributed pairs of spins, with soft ferromagnetic coupling $J$= 0.1 $J_{FM}$ (§2.2) [31] and a randomly oriented anisotropy axes, with $K$= 4 $J_{FM}$.

Since the study entails NPs with cubic morphology (C12), we have also simulated the M-H loops, for different applied $H_{cool}$, of defected cubic NPs by substituting the uniaxial, with cubic core anisotropy. Neither the $H_C$ nor the $H_{EB}$ appear to deviate from the originally obtained behaviour with uniaxial anisotropy (Fig. S17).